\documentclass[conference]{IEEEtran}
\usepackage{graphicx}         
\usepackage{graphicx}
\usepackage {subfigure}     
\usepackage{epstopdf}
\usepackage{subfloat}

\usepackage[table]{xcolor}
\usepackage{color}
\begin{document}
%
\title{Radiation Pattern Synthesis Using Hybrid Fourier- Woodward-Lawson-Neural Networks for Reliable MIMO Antenna Systems} 


\author{\IEEEauthorblockN{Elies Ghayoula$^{1,2}$, Ridha Ghayoula$^2$, Jaouhar Fattahi$^3$, Emil Pricop$^4$,\\ Jean-Yves Chouinard$^2$ and Ammar Bouallegue$^1$} 
\IEEEauthorblockA{$^1$Sys'Com Laboratory, National Engineering School of Tunis, ENIT, Tunis El Manar University, Tunisia.}

\IEEEauthorblockA{$^2$Department of Electrical and Computer Engineering, Laval University, Quebec, Canada.}

\IEEEauthorblockA{$^3$Department of Computer Science and Software Engineering, Laval University, Quebec, Canada.}

\IEEEauthorblockA{$^4$Automatic Control, Computers and Electronics Department Petroleum-Gas University of Ploiesti, Romania.}

%
}


%
%

 
\maketitle

\begin{abstract}

In this paper, we implement hybrid Woodward-Lawson-Neural Networks and weighted Fourier method to synthesize antenna arrays. The neural networks (NN) is applied here to simplify the modeling of MIMO antenna arrays by assessing phases. The main problem is obviously to find optimal weights of the linear antenna array elements giving radiation pattern with minimum sidelobe level (SLL) and hence ameliorating the antenna array performance. To attain this purpose, an antenna array for reliable Multiple-Input Multiple-Output (MIMO) applications with frequency at 2.45 GHz is implemented. To validate the suggested method, many examples of uniformly excited array patterns with the main beam are put in the direction of the useful signal. The Woodward-Lawson-Neural Networks synthesis method permits to find out interesting analytical equations for the synthesis of an antenna array and highlights the flexibility between the system parameters in input and those in output. The performance of this hybrid optimization underlines how well the system is suitable for a wireless communication and how it participates in reducing interference, as well.

\end{abstract}


%
\IEEEpeerreviewmaketitle

\section{Introduction} 

Antenna array are largely used in applications including satellites, wireless communications, MIMO systems, remote detection, radars, biomedical imaging, and so on. Pattern synthesis consists in choosing the antenna aspects such as the desired SLL, the most suitable position of the nulls, as well as the beamwidth of antenna pattern, and this is in order to reach the  radiation pattern near to the optimal one. In literature, Several beamforming techniques exist. For instance, a synthesis of radiation patterns in linear arrays based on genetic algorithms (GA) is shown in \cite{art1}. In \cite{art2} the authors exhibit an SLL reduction in linear array pattern synthesis based on particle swarm optimization (PSO). In the same vein, a linear array thinning based on iterative Fourier techniques is presented in \cite{art6}. A phase-only synthesis has also been presented in \cite{art7} to attain the most suitable radiation pattern. The criterion of nonlinearity related to antenna radiation patterns make antennas appropriate to artificial neural networks (ANN). NNs have been proven good enough to cope with problems when the link between inputs and outputs is cumbersome and hard to formulate. They can reasonably assess a model owing to their capabilities to adapt their parameters by utilizing known input/output pairs. They are also able to provide optimal weights between neurons after receiving a training phase. As soon as the training phase is finished, the network ready to communicate the results to the inputs. The increasing utilization of ANNs in a plethora of electromagnetic applications motivated its use in antenna array design in radiation optimization and synthesis contexts. The Multilayer Perceptron (MLP) is one of the oldest model of NN that implements single hidden layer. This kind of network is capable of approximating a given smooth nonlinear input-output mapping to an arbitrary degree of precision if a sufficient set of hidden layer neurons is provided \cite{art10}. For instance, a phased array using Taguchi-neural networks is given in \cite{art12}. In \cite{art20} the authors’ present a usual application of back-propagation NN for synthesis and optimization of antenna array. In this paper, we are interested in presenting an adaptive Fourier Woodward-Lawson-Neural networks technique to synthesize of MIMO smart antennas. A contribution regarding several features is here made. The paper is organized as follows: The formulation of the synthesis problem is described in section II. Fourier Woodward-Lawson method is described in section III. ANN is implemented by simulation in section IV. Validation using CST microwave studio for linear MIMO antennas array with 16 elements is provided in section V. Finally, in section VI we make some overall conclusions.

\section{PROBLEM FORMULATION}
\begin{figure} [!ht]
\centerline{\includegraphics[width=0.25\textwidth,draft=false]{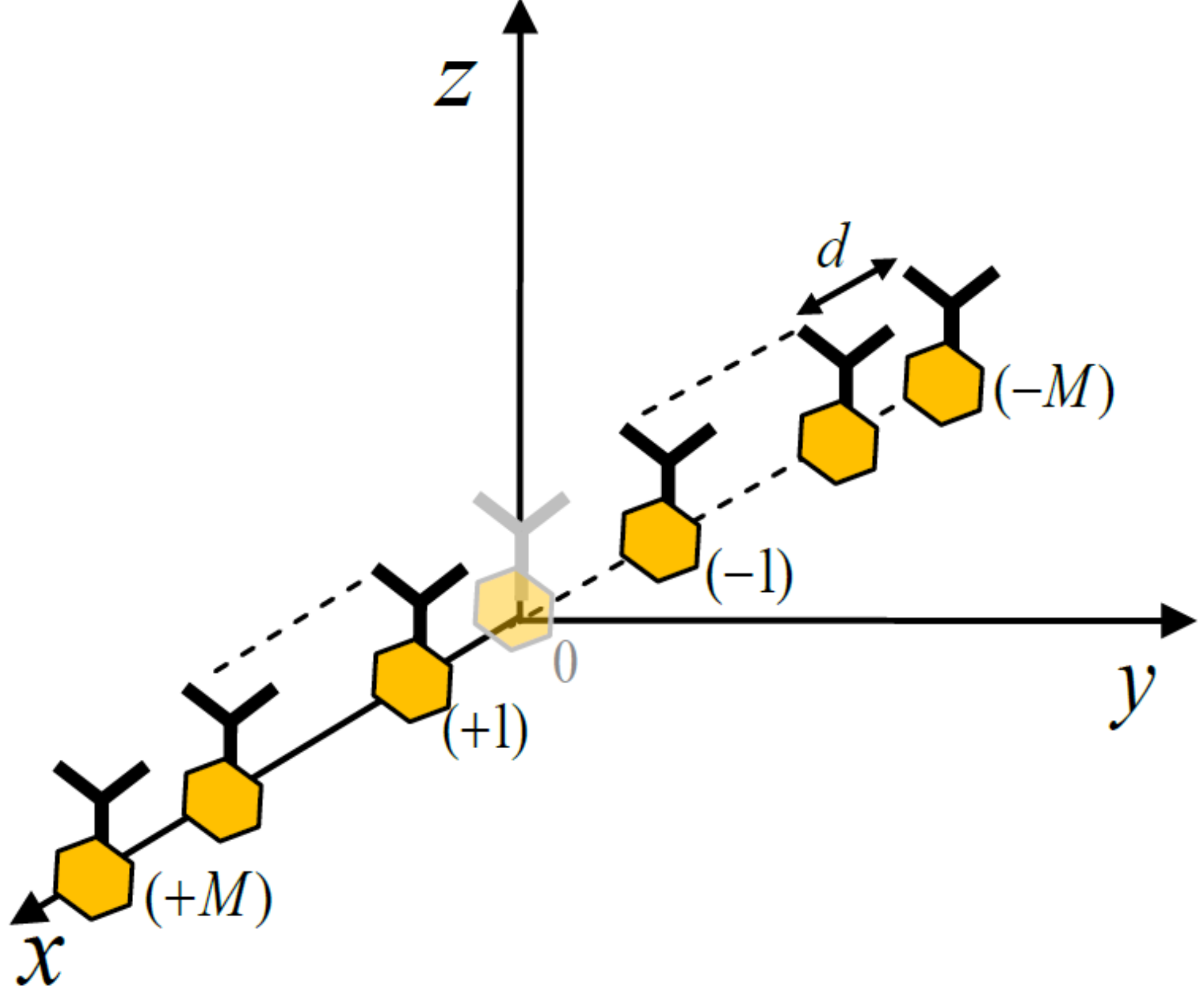}}
\caption{Linear antenna array.}
\label{fig1}
\end{figure}

Our model as shwon in Fig. \ref{fig1} is a linear antenna array with $N$ isotropic elements placed symmetrically along the x-axis.
The far-field $F\left( u \right)$ of this model below with $N$ antennas arranged along a periodic grid at distance $d$, can be expressed as the product of the array factor $AF\left( u \right)$ and the embedded element pattern $EF\left( u \right)$    

\begin{equation}
F\left( u \right) = AF\left( u \right)EF\left( u \right)
\label{equ1}
\end{equation}

\begin{equation}
AF\left( u \right) = \sum\limits_{n = 0}^{N - 1} {w_n } e^{jkndu}
\label{equ2}
\end{equation}
where {$w_n$ is the complex weighting of the $n^{th}$ element of our linear array}, {$k$ is the wavenumber $\left( {\frac{{2\pi }}{\lambda }} \right)$}, {$\lambda $ is the wavelength} and {$u = \cos \theta$ where $\theta$ is the angular coordinate between far-field direction and the normal array model.} Here, the problem of synthesis to be solved is to find the optimal weights (phases and amplitudes) of our model that are able to perform the radiation in desired direction by controlling the phase of each antenna with maximum SLL reduction by controlling the amplitude.

\section{Hybrid Fourier-Woodward Lawson method for Frequency-Sampling Design}

The array factor with the Discrete Fourier Transform (DFT) is described with details in \cite{article27}. To justify the use of Fourier approach with Woodward-Lawson method, many popular optimization algorithms are used for comparison; such as Schelkunoff \cite{article26}, Dolph-Chebyshev  and Taylor \cite{article24}. Here, Fourier methods give us the best results. 

\begin{figure} [!ht]
\centering{
\includegraphics[width=0.32\textwidth]{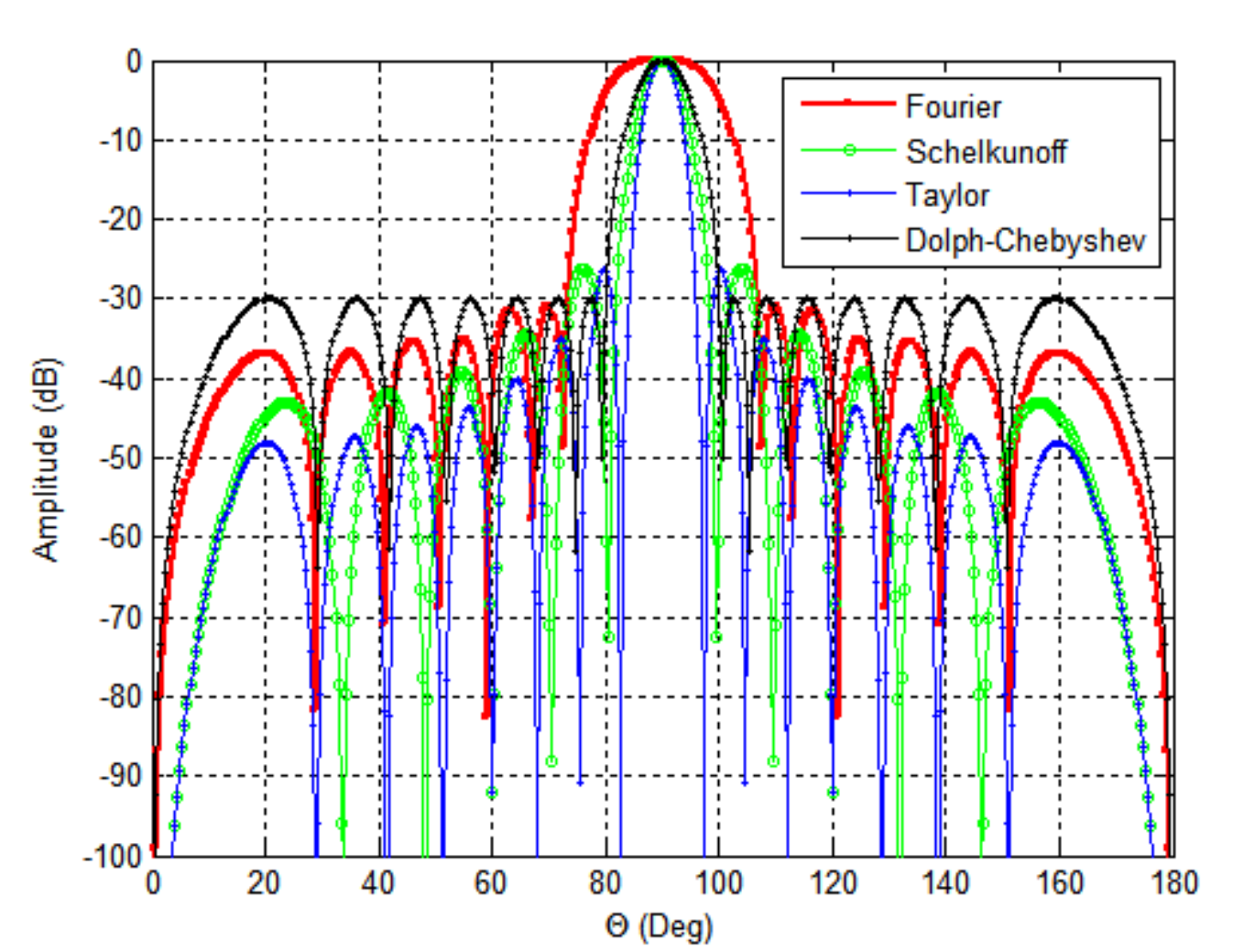}
\caption{Comparison of optimization algorithms for our model.}
\label{comparison}}
\end{figure}

Fig. \ref{comparison} shows the radiation patterns of Fourier method compared to Schelkunoff, Taylor and Dolph-Chebyshev method. It's clear from results Fig. \ref{comparison} that SLL here is reduced to $-30$ dB and more for all optimization techniques with different values of half power beamwidth (HPBW) about $7^\circ$ with Schelkunoff, $4.5^\circ$ with Taylor, $8^\circ$ with Dolph-Chebyshev and about $18^\circ$ with Fourier. For our model with $16$ antennas, optimal weights generated using Fourier method are shown in TABLE \ref{tab1}. all these results confirm that radiation patterns synthesis with Fourier outperform the other techniques.

\begin{table*}[ht]
\caption{Optimal excitations with Fourier for linear antenna array ($M=16$ and  $d = 0.5\lambda$).}
\begin{center}
\renewcommand{\arraystretch}{1.05}
\small{
\begin{tabular}{|l|l|l|l|l|l|l|l|l|l|l|l|}
\cline{2-12}
\hline
\multicolumn{1}{|c|}{\textbf{Number of}} & \multicolumn{11}{c|}{\textbf{Phase Vs Amplitude}} \\ 
\cline{2-12}
\multicolumn{1}{|c|}{\textbf{elements m}} & \multicolumn{1}{c|}{\textbf{40$^\circ$}} & \multicolumn{1}{c|}{\textbf{50$^\circ$}} & \multicolumn{1}{c|}{\textbf{60$^\circ$}} & \multicolumn{1}{c|}{\textbf{70$^\circ$}} & \multicolumn{1}{c|}{\textbf{80$^\circ$}} & \multicolumn{1}{c|}{\textbf{90$^\circ$}} & \multicolumn{1}{c|}{\textbf{100$^\circ$}} & \multicolumn{1}{c|}{\textbf{110$^\circ$}} & \multicolumn{1}{c|}{\textbf{120$^\circ$}} & \multicolumn{1}{c|}{\textbf{130$^\circ$}} & \multicolumn{1}{c|}{\textbf{140$^\circ$}} \\ 
\hline
\multicolumn{1}{|c|}{\textbf{1\&16}} & \multicolumn{1}{c|}{0.0095} & \multicolumn{1}{c|}{0.0128} & \multicolumn{1}{c|}{0.0149} & \multicolumn{1}{c|}{0.0160} & \multicolumn{1}{c|}{0.0165} & \multicolumn{1}{c|}{0.0166} & \multicolumn{1}{c|}{0.0165} & \multicolumn{1}{c|}{0.0160} & \multicolumn{1}{c|}{0.0149} & \multicolumn{1}{c|}{0.0128} & \multicolumn{1}{c|}{0.0095} \\ 
\hline
\multicolumn{1}{|c|}{\textbf{2\&15}} & \multicolumn{1}{c|}{0.0023} & \multicolumn{1}{c|}{0.0078} & \multicolumn{1}{c|}{0.0119} & \multicolumn{1}{c|}{0.0147} & \multicolumn{1}{c|}{0.0163} & \multicolumn{1}{c|}{0.0169} & \multicolumn{1}{c|}{0.0163} & \multicolumn{1}{c|}{0.0147} & \multicolumn{1}{c|}{0.0119} & \multicolumn{1}{c|}{0.0078} & \multicolumn{1}{c|}{0.0023} \\ 
\hline
\multicolumn{1}{|c|}{\textbf{3\&14}} & \multicolumn{1}{c|}{0.0143} & \multicolumn{1}{c|}{0.0079} & \multicolumn{1}{c|}{0.0025} & \multicolumn{1}{c|}{0.0015} & \multicolumn{1}{c|}{0.0040} & \multicolumn{1}{c|}{0.0048} & \multicolumn{1}{c|}{0.0040} & \multicolumn{1}{c|}{0.0015} & \multicolumn{1}{c|}{0.0025} & \multicolumn{1}{c|}{0.0079} & \multicolumn{1}{c|}{0.0143} \\ 
\hline
\multicolumn{1}{|c|}{\textbf{4\&13}} & \multicolumn{1}{c|}{0.0405} & \multicolumn{1}{c|}{0.0352} & \multicolumn{1}{c|}{0.0303} & \multicolumn{1}{c|}{0.0264} & \multicolumn{1}{c|}{0.0239} & \multicolumn{1}{c|}{0.0230} & \multicolumn{1}{c|}{0.0239} & \multicolumn{1}{c|}{0.0264} & \multicolumn{1}{c|}{0.0303} & \multicolumn{1}{c|}{0.0352} & \multicolumn{1}{c|}{0.0405} \\ 
\hline
\multicolumn{1}{|c|}{\textbf{5\&12}} & \multicolumn{1}{c|}{0.0736} & \multicolumn{1}{c|}{0.0716} & \multicolumn{1}{c|}{0.0693} & \multicolumn{1}{c|}{0.0671} & \multicolumn{1}{c|}{0.0656} & \multicolumn{1}{c|}{0.0651} & \multicolumn{1}{c|}{0.0656} & \multicolumn{1}{c|}{0.0671} & \multicolumn{1}{c|}{0.0693} & \multicolumn{1}{c|}{0.0716} & \multicolumn{1}{c|}{0.0736} \\ 
\hline
\multicolumn{1}{|c|}{\textbf{6\&11}} & \multicolumn{1}{c|}{0.1082} & \multicolumn{1}{c|}{0.1109} & \multicolumn{1}{c|}{0.1125} & \multicolumn{1}{c|}{0.1133} & \multicolumn{1}{c|}{0.1137} & \multicolumn{1}{c|}{0.1138} & \multicolumn{1}{c|}{0.1137} & \multicolumn{1}{c|}{0.1133} & \multicolumn{1}{c|}{0.1125} & \multicolumn{1}{c|}{0.1109} & \multicolumn{1}{c|}{0.1082} \\ 
\hline
\multicolumn{1}{|c|}{\textbf{7\&10}} & \multicolumn{1}{c|}{0.1374} & \multicolumn{1}{c|}{0.1447} & \multicolumn{1}{c|}{0.1504} & \multicolumn{1}{c|}{0.1544} & \multicolumn{1}{c|}{0.1568} & \multicolumn{1}{c|}{0.1576} & \multicolumn{1}{c|}{0.1568} & \multicolumn{1}{c|}{0.1544} & \multicolumn{1}{c|}{0.1504} & \multicolumn{1}{c|}{0.1447} & \multicolumn{1}{c|}{0.1374} \\ 
\hline
\multicolumn{1}{|c|}{\textbf{8\&9}} & \multicolumn{1}{c|}{0.1549} & \multicolumn{1}{c|}{0.1652} & \multicolumn{1}{c|}{0.1735} & \multicolumn{1}{c|}{0.1796} & \multicolumn{1}{c|}{0.1833} & \multicolumn{1}{c|}{0.1845} & \multicolumn{1}{c|}{0.1833} & \multicolumn{1}{c|}{0.1796} & \multicolumn{1}{c|}{0.1735} & \multicolumn{1}{c|}{0.1652} & \multicolumn{1}{c|}{0.1549} \\ 
\hline
\end{tabular}

}
\label{tab1}
\end{center}
\end{table*}

The Woodward-Lawson synthesis method \cite{WWL3} is based on the Fourier-transform relationship between the far-field pattern and the field in a planar aperture. This method is described in many classic books on antennas \cite{WWL6}. The array factor can be written as shown below:
\begin{equation}
AF = \sum\limits_{n = 1}^N {w_n e^{j\left( {n - 1} \right)\psi } }; \psi  = kd\cos \theta  + \beta  
\end{equation}
where $w_n$, $\beta$ and $d$ represent amplitude, phase and position of the $n^{th}$ element respect to the origin.

The radiation characteristics of continuous sources can be approximated by discrete-element arrays:
\begin{equation}
w_n e^{j\left( {n - 1} \right)\beta }  = I_n (z)e^{j\phi _n \left( z \right)}
\end{equation}

Let the source be represented by a sum of the following constant current source of length $l$.

\begin{equation}
i_m \left( z \right) = \frac{{b_m }}{l}e^{ - jkz\cos \theta _m } ; - l/2 \le z \le l/2
\end{equation}

Then the current source can be given by

\begin{equation}
I\left( z \right) = \frac{1}{l}\sum\limits_{m =  - M}^M {b_m } e^{ - jkz\cos \theta _m } 
\end{equation}
where $m =  \pm 1, \pm 2, \ldots , \pm M$ (for $2M$ even number) and $m =  0, \pm 1, \pm 2, \ldots , \pm M$ (for $2M+1$ odd number) 
\begin{figure} [!ht]
\centering{
\includegraphics[width=0.32\textwidth]{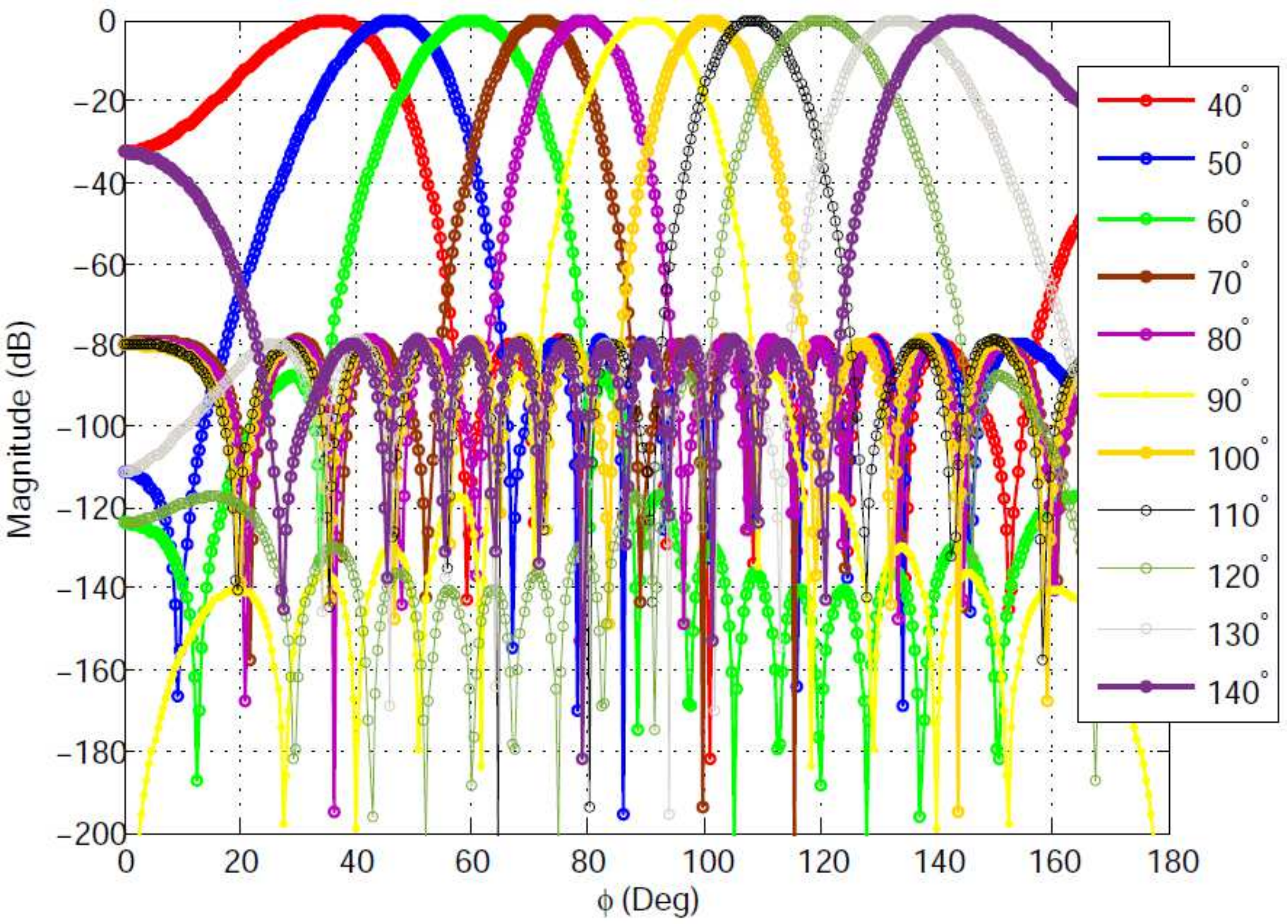}
\caption{Space scanning for our model using Woodward-Lawson method with Fourier technique.}
\label{phase_matlab}
}
\end{figure}

Applying Woodward-Lawson means sampling the desired pattern at various discrete locations. this method use composing function of the forms: $\frac{{b_m \sin \left( {\psi _m } \right)}}{{\psi _m }}$ or $\frac{{b_m \sin \left( {N\phi _m } \right)}}{{N\sin \phi _m }}$ The synthesized pattern is represented by a finite sum of composing functions. The total excitation is a sum of space harmonics. 17 desired direction of uniform linear antenna array with $N=16$ isotropic elements were selected in order to illustrate the performance of the proposed method for space scanning beams in desired direction by controlling the phase excitation of each array element. By using Woodward-Lawson method with Fourier technique in different sectors, the computational results from Fig. \ref{phase_matlab} show a high performance phase control capability for beam pattern synthesis.

\section{Neural Networks}
\begin{figure} [!ht]
\centering{
\includegraphics[width=0.32\textwidth]{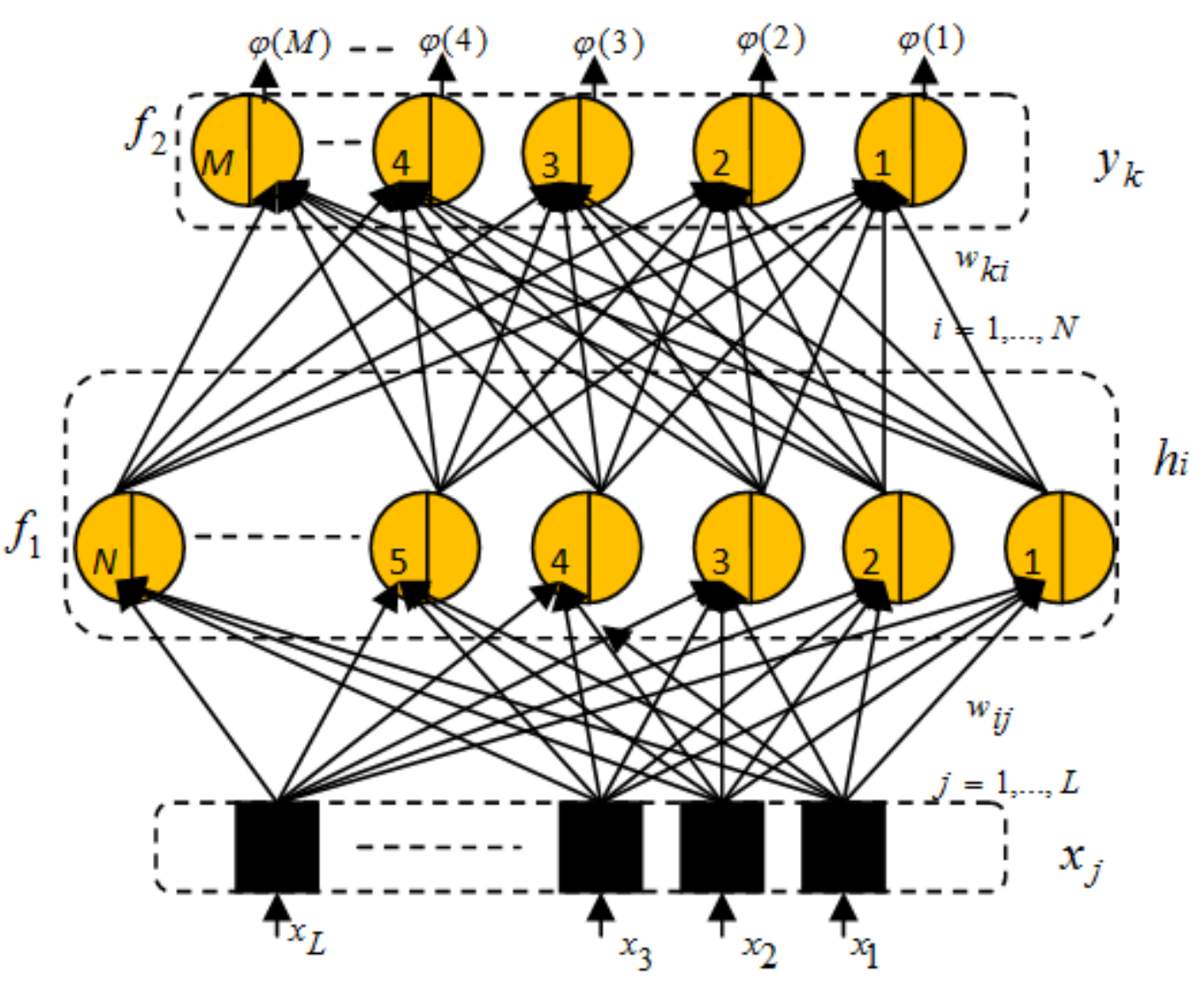}
\caption{Architecture of neural beamformer.}
\label{NN1}}
\end{figure}

ANN is applied here to approximate the input and output relationship by optimizing the weights using known input-output training pairs. Once the training is complete, it is able to obtain the requested radiation beam of our antenna for a given set of inputs by adaptive signal processing \cite{NN1}. The multi-layer networks are composed of an input layer whose neurons encode the information given to the network, a variable number of internal layers called "hidden" and an output layer containing as many neurons as the desired responses (Fig. \ref{NN1}). The training process of these networks is supervised. The first layer is composed of input nodes. A MLP network is an advanced NN stream with a hidden layer, with a multi-layer perceptron node function at each hidden node. The number of nodes $L$ is equal to the dimension of input vector \cite{a17}. From Fig. \ref{NN1} we denote the index of hidden layer by $i$ with $i = 1,2,...,N$, the index of input layer by $i$ with $j = 1,2,...,L$ and the index of output layer by $k$ with $k = 1,2,...,M$. The interconnection weights are generated based on the requirement of minimum error between the training data $d_k$ and the neural model output $y_k$. The purpose of the training procedure is to adapt the network interconnection weights $w_{ij}$ and $w_{ki}$ in order to minimize the following error function.

\begin{equation}
E\left( t \right) = \frac{1}{2}\sum\limits_{k = 1}^M {\sum\limits_{i = 1}^N {\sum\limits_{j = 1}^L {\left[ {y_k \left( {x_j ,w_{ij} ,w_{ki} } \right) - d_k } \right]^2 } } } 
\label{equ13}
\end{equation}
where $t = 1,2,...,T$  is the index of the training set. This is an iterative process using the back-propagation algorithm described in \cite{a12}. For each iteration, the weights $w_{ij}$ and $w_{ki}$ are updated by

\begin{equation}
\Delta w_m  =  - \eta \frac{{\partial E}}{{\partial w_m }}
\label{equ14}
\end{equation}

\begin{figure} [!ht] 
\centering{
\includegraphics[width=0.4\textwidth]{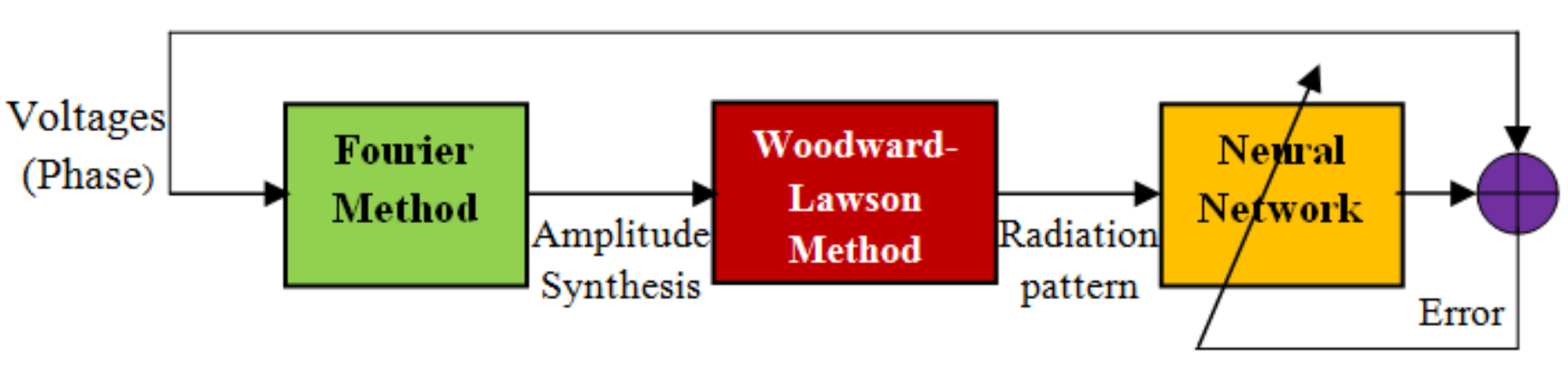}
\caption{Training procedure for NN.}
\label{NN2}}
\end{figure}

Fig. \ref{NN2} illustrates a block diagram of the mean square error for MLP Network.

\begin{figure*}[!ht] 
    \centering 
    \subfigure[Training performance]{\label{NN3}\includegraphics[width=5.5cm]{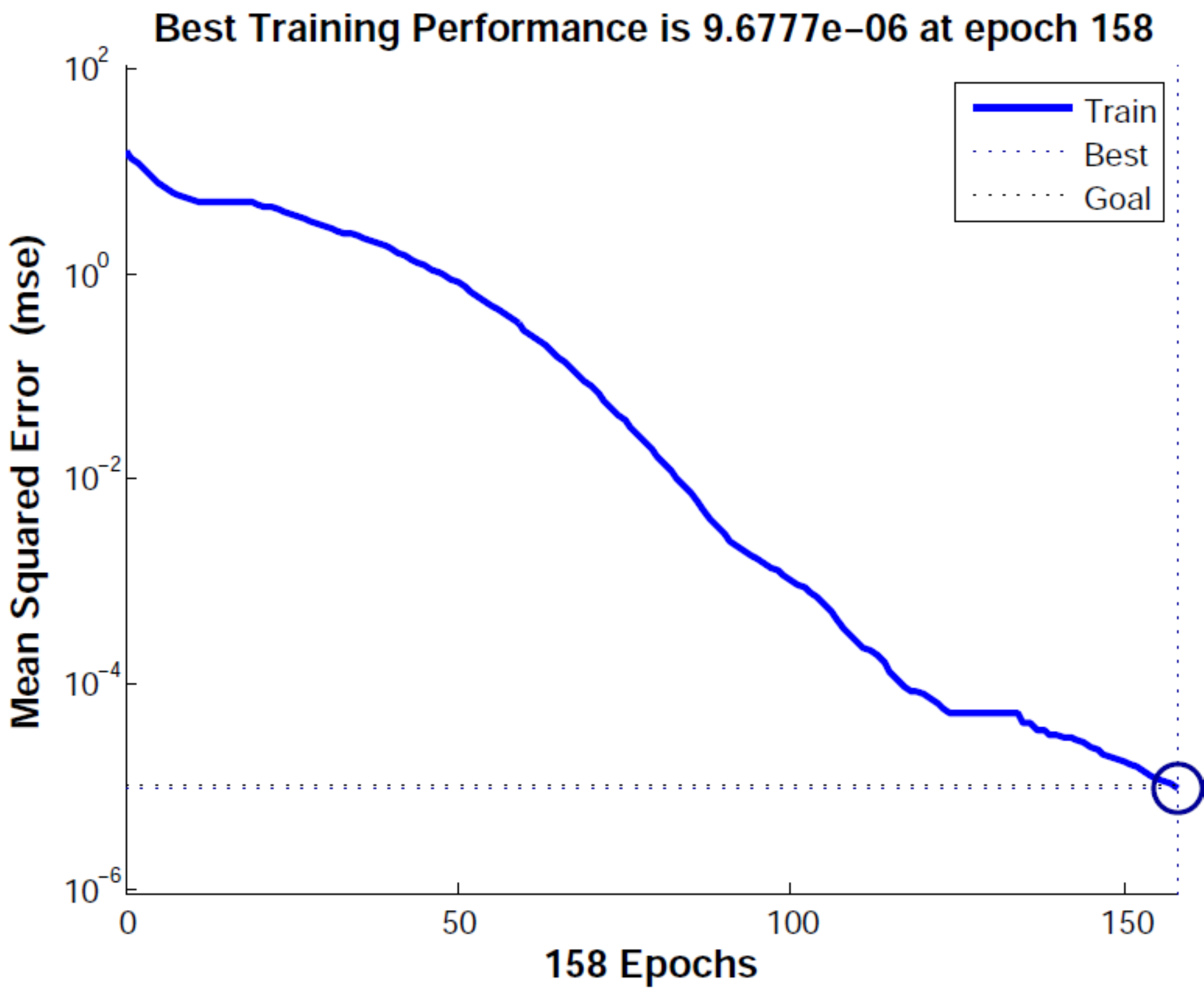}} 
    \subfigure[NN performance]{\label{NN4}\includegraphics[width=6cm]{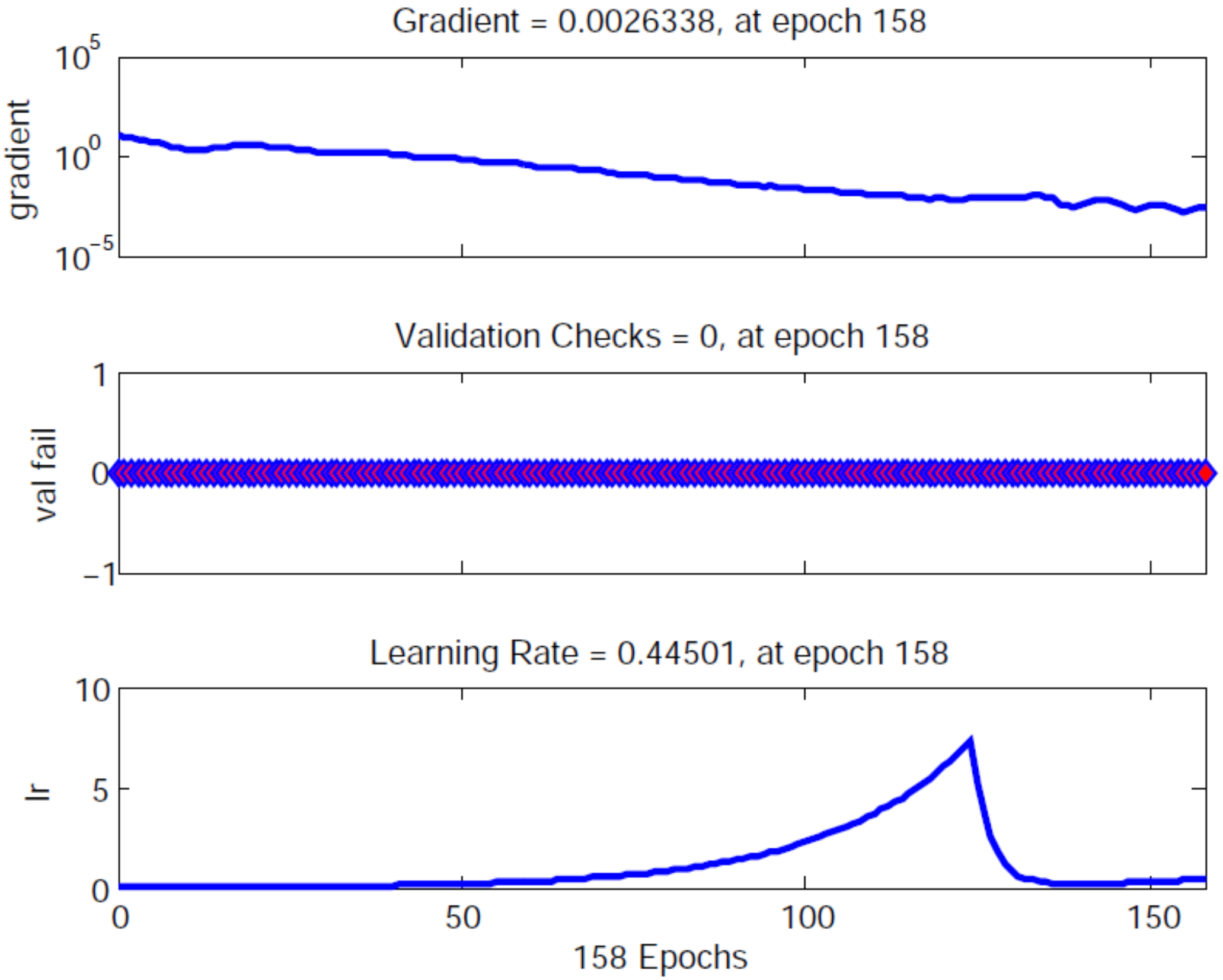}} 
		\subfigure[Network created during training]{\label{NN5}\includegraphics[width=5cm]{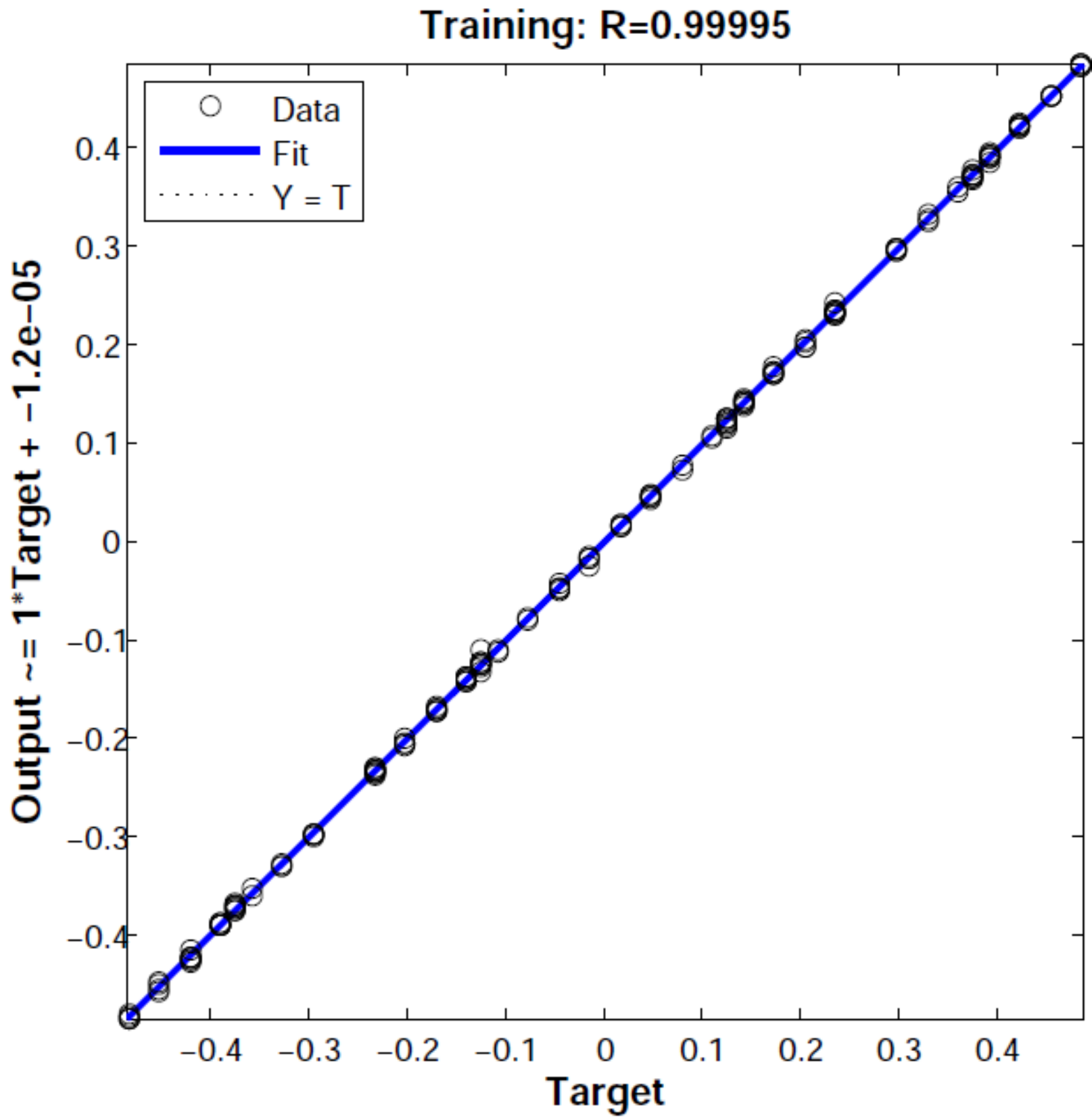}} 
    \caption{Neural networks results.} 
    \label{NN3_4_5} 
\end{figure*}

The ability of generalize is one of the main advantages of NN. it means that a trained network able to classify data from the same category as the learning data that it has never seen before. In real life applications, developers have only a small part of all possible radiation patterns for the generation of a NN. To achieve the best generalization, the dataset should be divided into three parts: The training set which is used to train a NN; the error of this dataset which is minimized during training, The validation set which is used then to define the performance of a NN on patterns that are not trained during learning phase and a test set for finally checking the total performance of a NN. We have two steps:
\begin{enumerate}
\item [1.] Designing network: Start with forming the input vectors $\left\{ {x_p ,p = 1,2,...,16} \right\}$, after that generate input/output pairs $\left\{ {x_p ,\varphi _q } \right\}$, where $q = 1,2,...,18$ and finally design the NN.
\item [2.] Testing network (Generalization): Start with forming the vectors $x_p^{'}$ for testing the input samples, after that prepare input vectors $x_p^{'}$ to the NN and at the end, get the output of the network.
\end{enumerate}

The choice of the number of hidden neurons depends to the nature of nonlinearity of model. Here, in our case in Table \ref{tab2}, the number of hidden neurons allowed a rapid convergence of the algorithm is $30$ and given a good precision of the formed neuronal model. The neuron used in this model here, is the continuous nonlinear neuron with tan sigmoid as activation function.

\begin{table}[!ht]
\caption{Back-Propagation algorithm.}
\begin{center}
\renewcommand{\arraystretch}{1}
\small{
\begin{tabular}{|l|l|l|}
\hline
\multicolumn{1}{|c|}{\textbf{Parameters}} & \multicolumn{1}{c|}{\textbf{Symbol}} & \multicolumn{1}{c|}{\textbf{Value}} \\ 
\hline
\multicolumn{1}{|c|}{Training coefficient } & \multicolumn{1}{c|}{$\eta$} & \multicolumn{1}{c|}{0.02} \\ 
\hline
\multicolumn{1}{|c|}{Neuron for input layer} & \multicolumn{1}{c|}{$n$} & \multicolumn{1}{c|}{18} \\ 
\hline
\multicolumn{1}{|c|}{Neuron for output layer} & \multicolumn{1}{c|}{$m$} & \multicolumn{1}{c|}{16} \\ 
\hline
\multicolumn{1}{|c|}{Neuron for hidden layer} & \multicolumn{1}{c|}{$h$} & \multicolumn{1}{c|}{30} \\ 
\hline

\end{tabular}

}
\label{tab2}
\end{center}
\end{table}


A linear antenna array with $16$ elements is now implemented for radiation pattern synthesis with the same amplitude and variable phase \cite{a17}. Simulation results must show radiation patterns with maximum SLL reduction at least $-20$dB and maintaining main lobes in the direction of useful signal for the reference antenna array ($16$ antennas). In our real application, the desired radiation pattern synthesis must be between $+40^\circ$ and $+140^\circ$, the database of our model contains whole data(input/output) generated using Fourier optimization. Fig. \ref{NN3_4_5} (\ref{NN3}, \ref{NN4} and \ref{NN5}) illustrates all phases of training, validation, performance and testing data of the proposed NN model. Fig. \ref{NN5} represents the graphical output generated by regression. The network outputs are plotted versus the targets as open circles. The perfect fit (output equal to targets) is indicated by the solid line. The best linear fit is indicated by a dashed line. In this application, because of the fit is so good, it is very difficult to choose between the best and the perfect linear fit line.

By using NN with Fourier technique in many sectors, $10$ desired direction of uniform linear antenna array with N = 16 isotropic elements were selected in order to illustrate the performance of the proposed method for space scanning beams in desired direction by controlling the phase excitation of each array element. simulation results prove an excellent phase control capability for beam pattern synthesis.

\section{Design and evaluation of MIMO Antennas with CST Microwave Studio}

\begin{figure} [!ht]
\centering{
\includegraphics[width=0.35\textwidth]{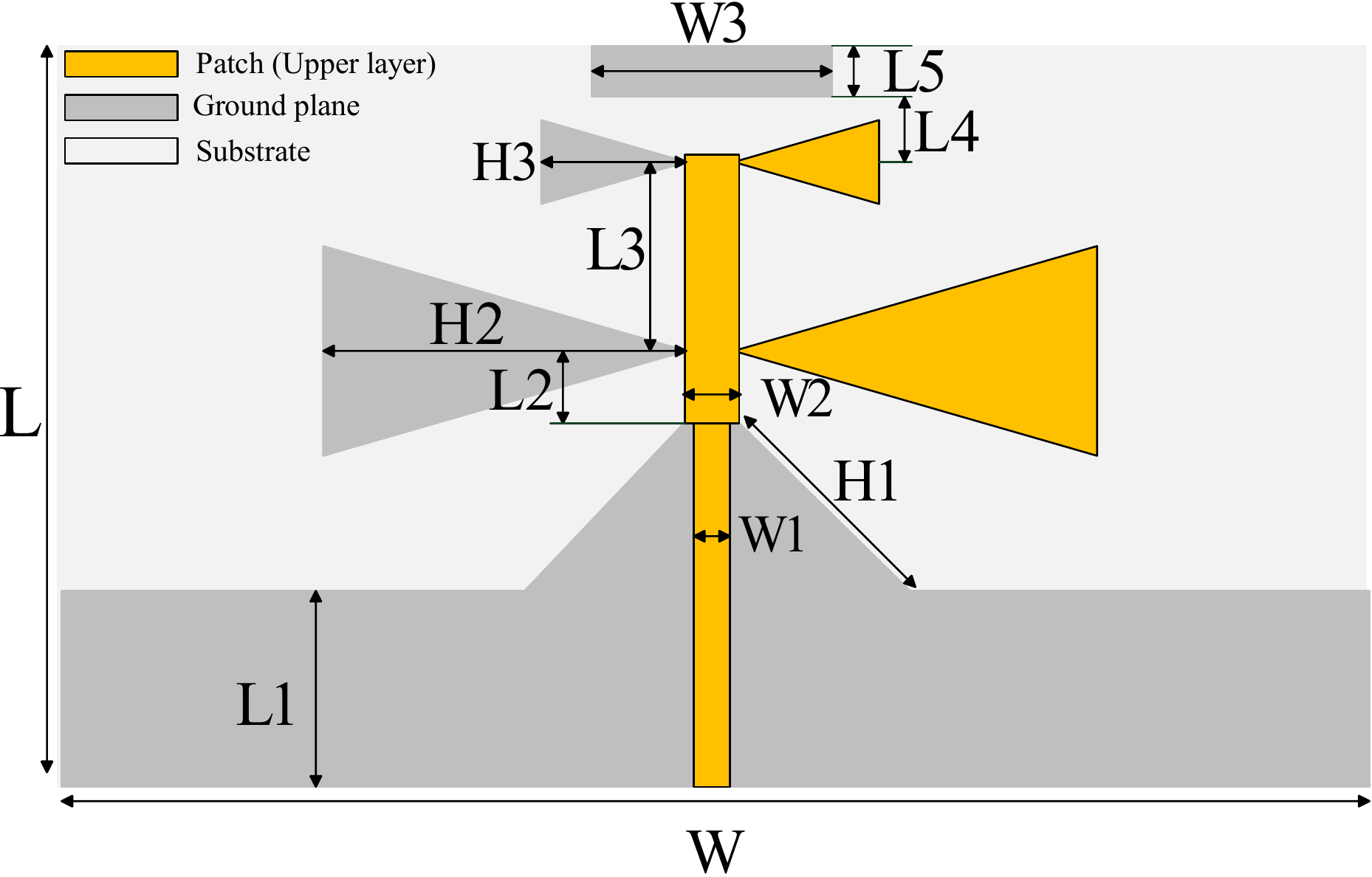}
\caption{Geometry of the proposed antenna.}
\label{antenna}}
\end{figure}

\begin{figure}[!ht] 
    \centering 
    \subfigure[Reflection coefficient.]{\label{coef}\includegraphics[width=4.4cm]{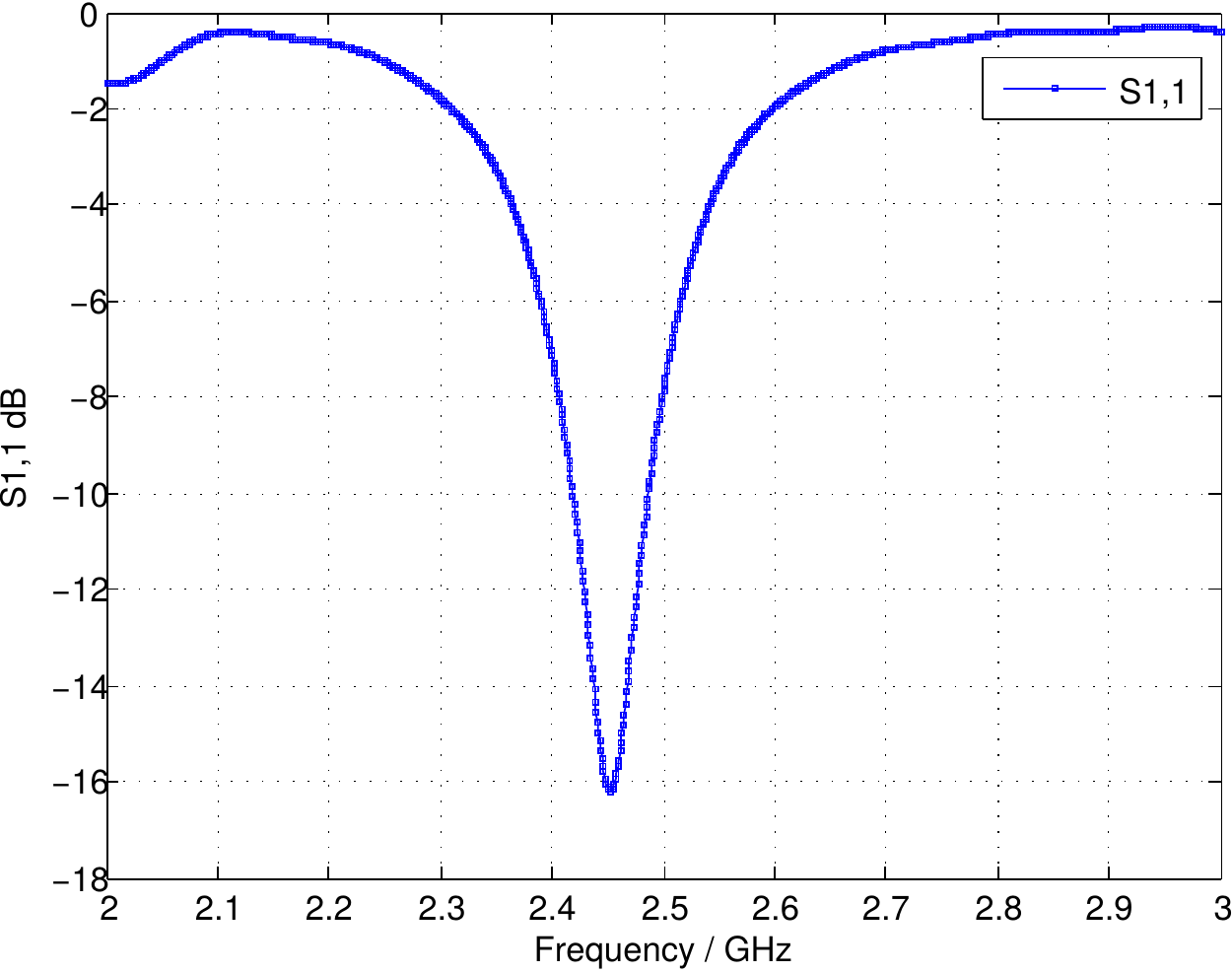}} 
    \subfigure[3D antenna radiation pattern.]{\label{diag_ray2}\includegraphics[width=4.3cm]{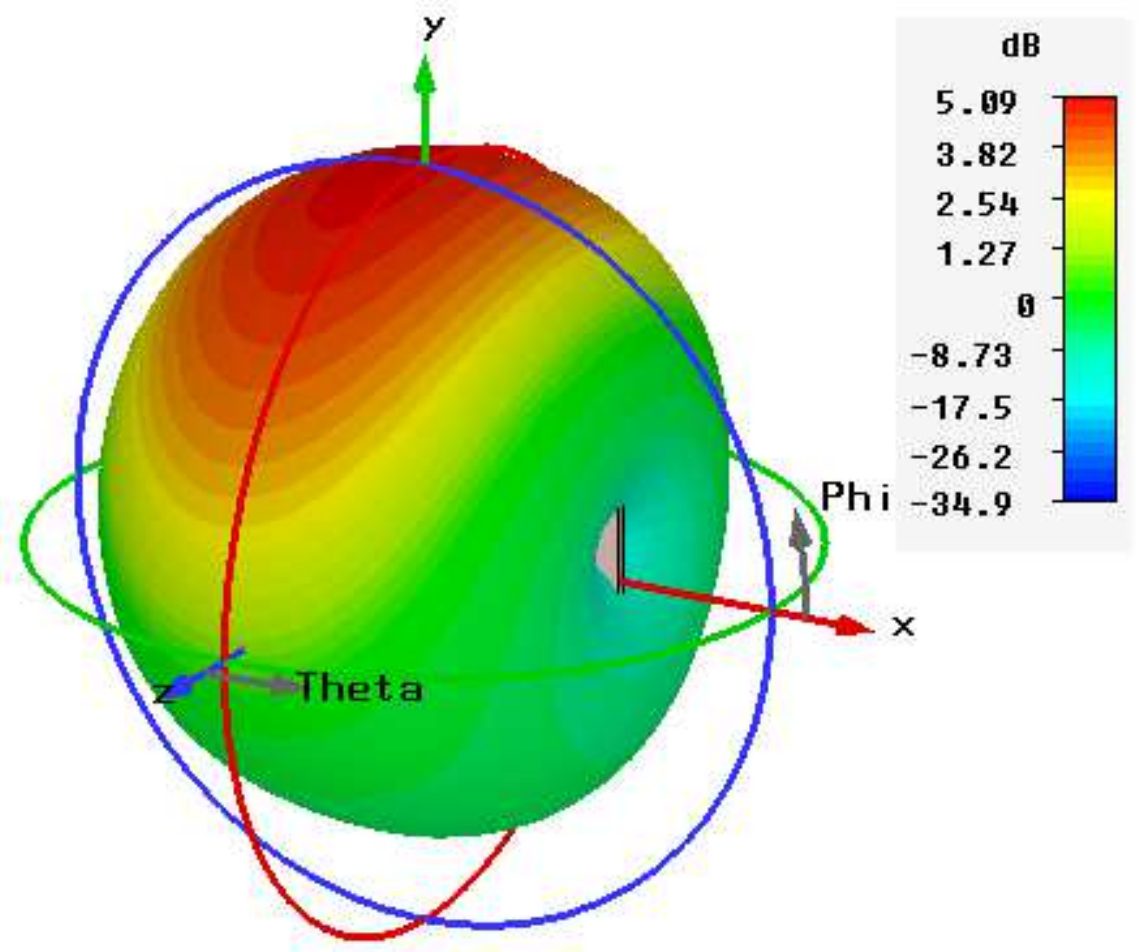}}  
    \caption{Simulation results at $2.45$ GHz.} 
    \label{fig4} 
\end{figure}

The proposed patch antenna shown in Fig. \ref{antenna} is designed on a $1mm$ thick FR-4 substrate for 2.45 GHz and 5 GHz WLAN. This antenna is similar to the antenna presented in \cite{bowtie}. The dimension of the antenna is $26 \times 60$ mm. $W1$, $W2$, $W3$, $R1$, $R2$, $R3$, $L1$, $L2$, $L3$, $L4$, and $L5$ are $2.50$, $2.46$, $12.5$, $6$, $16.3$, $7.4$, $5$, $2$, $7$, $3$, and $3$ mm, respectively. Fig. \ref{coef} shows the measured reflection coefficients of the antenna, which fully covers the 2.45-GHz band used for MIMO application \cite{article15}. 

\begin{table*}[!ht]
\caption{Optimal excitations with Woodward-Lawson-Neural Networks (WWL-NN) for linear antenna array ($M=16$ and  $d = 0.5\lambda$).}
\begin{center}
\renewcommand{\arraystretch}{1}
\small{
\begin{tabular}{|l|l|l|l|l|l|l|l|l|l|l|}
\cline{2-11}
\hline
\multicolumn{1}{|c|}{\textbf{Number of} } & \multicolumn{10}{c|}{\textbf{Phase(deg.)}} \\ 
\cline{2-11}
\multicolumn{1}{|c|}{\textbf{elements m}} & \multicolumn{1}{c|}{\textbf{40$^\circ$}} & \multicolumn{1}{c|}{\textbf{50$^\circ$}} & \multicolumn{1}{c|}{\textbf{60$^\circ$}} & \multicolumn{1}{c|}{\textbf{70$^\circ$}} & \multicolumn{1}{c|}{\textbf{80$^\circ$}} & \multicolumn{1}{c|}{\textbf{100$^\circ$}} & \multicolumn{1}{c|}{\textbf{110$^\circ$}} & \multicolumn{1}{c|}{\textbf{120$^\circ$}} & \multicolumn{1}{c|}{\textbf{130$^\circ$}} & \multicolumn{1}{c|}{\textbf{140$^\circ$}} \\ 
\hline
\multicolumn{1}{|c|}{\textbf{1}} & \multicolumn{1}{c|}{17.208} & \multicolumn{1}{c|}{-151.882} & \multicolumn{1}{c|}{-44.872} & \multicolumn{1}{c|}{61.224} & \multicolumn{1}{c|}{-107.736} & \multicolumn{1}{c|}{107.245} & \multicolumn{1}{c|}{-62.884} & \multicolumn{1}{c|}{44.222} & \multicolumn{1}{c|}{154.017} & \multicolumn{1}{c|}{-15.368} \\ 
\hline
\multicolumn{1}{|c|}{\textbf{2}} & \multicolumn{1}{c|}{-130.235} & \multicolumn{1}{c|}{83.084} & \multicolumn{1}{c|}{-135.415} & \multicolumn{1}{c|}{6.310} & \multicolumn{1}{c|}{-141.715} & \multicolumn{1}{c|}{139.125} & \multicolumn{1}{c|}{-7.446} & \multicolumn{1}{c|}{135.148} & \multicolumn{1}{c|}{-83.856} & \multicolumn{1}{c|}{132.842} \\ 
\hline
\multicolumn{1}{|c|}{\textbf{3}} & \multicolumn{1}{c|}{84.556} & \multicolumn{1}{c|}{-41.425} & \multicolumn{1}{c|}{133.411} & \multicolumn{1}{c|}{-49.156} & \multicolumn{1}{c|}{-176.033} & \multicolumn{1}{c|}{173.117} & \multicolumn{1}{c|}{49.4507} & \multicolumn{1}{c|}{-134.931} & \multicolumn{1}{c|}{40.874} & \multicolumn{1}{c|}{-82.394} \\ 
\hline
\multicolumn{1}{|c|}{\textbf{4}} & \multicolumn{1}{c|}{-62.037} & \multicolumn{1}{c|}{-163.176} & \multicolumn{1}{c|}{45.236} & \multicolumn{1}{c|}{-106.976} & \multicolumn{1}{c|}{151.753} & \multicolumn{1}{c|}{-151.587} & \multicolumn{1}{c|}{107.530} & \multicolumn{1}{c|}{-45.116} & \multicolumn{1}{c|}{162.731} & \multicolumn{1}{c|}{61.571} \\ 
\hline
\multicolumn{1}{|c|}{\textbf{5}} & \multicolumn{1}{c|}{151.836} & \multicolumn{1}{c|}{73.093} & \multicolumn{1}{c|}{-44.490} & \multicolumn{1}{c|}{-162.502} & \multicolumn{1}{c|}{116.645} & \multicolumn{1}{c|}{-118.579} & \multicolumn{1}{c|}{163.825} & \multicolumn{1}{c|}{44.591} & \multicolumn{1}{c|}{-72.056} & \multicolumn{1}{c|}{-151.937} \\ 
\hline
\multicolumn{1}{|c|}{\textbf{6}} & \multicolumn{1}{c|}{5.765} & \multicolumn{1}{c|}{-51.170} & \multicolumn{1}{c|}{-135.690} & \multicolumn{1}{c|}{141.227} & \multicolumn{1}{c|}{84.802} & \multicolumn{1}{c|}{-84.683} & \multicolumn{1}{c|}{-142.559} & \multicolumn{1}{c|}{135.067} & \multicolumn{1}{c|}{50.667} & \multicolumn{1}{c|}{-4.107} \\ 
\hline
\multicolumn{1}{|c|}{\textbf{7}} & \multicolumn{1}{c|}{-139.996} & \multicolumn{1}{c|}{-174.261} & \multicolumn{1}{c|}{134.156} & \multicolumn{1}{c|}{84.477} & \multicolumn{1}{c|}{50.758} & \multicolumn{1}{c|}{-50.330} & \multicolumn{1}{c|}{-82.326} & \multicolumn{1}{c|}{-135.194} & \multicolumn{1}{c|}{175.066} & \multicolumn{1}{c|}{140.313} \\ 
\hline
\multicolumn{1}{|c|}{\textbf{8}} & \multicolumn{1}{c|}{72.599} & \multicolumn{1}{c|}{63.637} & \multicolumn{1}{c|}{46.374} & \multicolumn{1}{c|}{28.325} & \multicolumn{1}{c|}{16.672} & \multicolumn{1}{c|}{-18.088} & \multicolumn{1}{c|}{-25.631} & \multicolumn{1}{c|}{-45.023} & \multicolumn{1}{c|}{-63.323} & \multicolumn{1}{c|}{-76.205} \\ 
\hline
\multicolumn{1}{|c|}{\textbf{9}} & \multicolumn{1}{c|}{-72.724} & \multicolumn{1}{c|}{-62.361} & \multicolumn{1}{c|}{-45.761} & \multicolumn{1}{c|}{-27.922} & \multicolumn{1}{c|}{-16.551} & \multicolumn{1}{c|}{17.140} & \multicolumn{1}{c|}{27.793} & \multicolumn{1}{c|}{44.645} & \multicolumn{1}{c|}{62.782} & \multicolumn{1}{c|}{73.722} \\ 
\hline
\multicolumn{1}{|c|}{\textbf{10}} & \multicolumn{1}{c|}{140.433} & \multicolumn{1}{c|}{173.796} & \multicolumn{1}{c|}{-135.864} & \multicolumn{1}{c|}{-86.155} & \multicolumn{1}{c|}{-50.513} & \multicolumn{1}{c|}{51.214} & \multicolumn{1}{c|}{84.942} & \multicolumn{1}{c|}{134.431} & \multicolumn{1}{c|}{-175.883} & \multicolumn{1}{c|}{-142.231} \\ 
\hline
\multicolumn{1}{|c|}{\textbf{11}} & \multicolumn{1}{c|}{-5.650} & \multicolumn{1}{c|}{50.282} & \multicolumn{1}{c|}{ 134.163} & \multicolumn{1}{c|}{-139.390} & \multicolumn{1}{c|}{-84.1103} & \multicolumn{1}{c|}{83.9749} & \multicolumn{1}{c|}{137.526} & \multicolumn{1}{c|}{-135.077} & \multicolumn{1}{c|}{-50.131} & \multicolumn{1}{c|}{7.566} \\ 
\hline
\multicolumn{1}{|c|}{\textbf{12}} & \multicolumn{1}{c|}{-152.083} & \multicolumn{1}{c|}{-73.762} & \multicolumn{1}{c|}{44.502} & \multicolumn{1}{c|}{162.706} & \multicolumn{1}{c|}{-121.094} & \multicolumn{1}{c|}{116.191} & \multicolumn{1}{c|}{-161.439} & \multicolumn{1}{c|}{-46.993} & \multicolumn{1}{c|}{75.012} & \multicolumn{1}{c|}{149.801} \\ 
\hline
\multicolumn{1}{|c|}{\textbf{13}} & \multicolumn{1}{c|}{62.369} & \multicolumn{1}{c|}{163.904} & \multicolumn{1}{c|}{-45.069} & \multicolumn{1}{c|}{106.216} & \multicolumn{1}{c|}{-151.567} & \multicolumn{1}{c|}{152.076} & \multicolumn{1}{c|}{-103.968} & \multicolumn{1}{c|}{44.138} & \multicolumn{1}{c|}{-163.161} & \multicolumn{1}{c|}{-65.510} \\ 
\hline
\multicolumn{1}{|c|}{\textbf{14}} & \multicolumn{1}{c|}{-85.610} & \multicolumn{1}{c|}{39.816} & \multicolumn{1}{c|}{-133.729} & \multicolumn{1}{c|}{50.269} & \multicolumn{1}{c|}{175.025} & \multicolumn{1}{c|}{-174.407} & \multicolumn{1}{c|}{-49.8276} & \multicolumn{1}{c|}{135.922} & \multicolumn{1}{c|}{-41.819} & \multicolumn{1}{c|}{84.289} \\ 
\hline
\multicolumn{1}{|c|}{\textbf{15}} & \multicolumn{1}{c|}{129.313} & \multicolumn{1}{c|}{-86.691} & \multicolumn{1}{c|}{133.341} & \multicolumn{1}{c|}{-3.937} & \multicolumn{1}{c|}{139.016} & \multicolumn{1}{c|}{-141.182} & \multicolumn{1}{c|}{4.562} & \multicolumn{1}{c|}{-135.199} & \multicolumn{1}{c|}{86.062} & \multicolumn{1}{c|}{-126.162} \\ 
\hline
\multicolumn{1}{|c|}{\textbf{16}} & \multicolumn{1}{c|}{-16.764} & \multicolumn{1}{c|}{151.248} & \multicolumn{1}{c|}{43.991} & \multicolumn{1}{c|}{-63.321} & \multicolumn{1}{c|}{106.209} & \multicolumn{1}{c|}{-106.390} & \multicolumn{1}{c|}{61.7133 } & \multicolumn{1}{c|}{-45.952} & \multicolumn{1}{c|}{-151.015} & \multicolumn{1}{c|}{16.507} \\ 
\hline
\end{tabular}

}
\label{tab3}
\end{center}
\end{table*}

\begin{figure*}[!ht] 
    \centering 
		\subfigure[ Perspective view of radiation synthesis at $40 ^\circ$]{\label{WLL_F_40_1}\includegraphics[width=4.75cm]{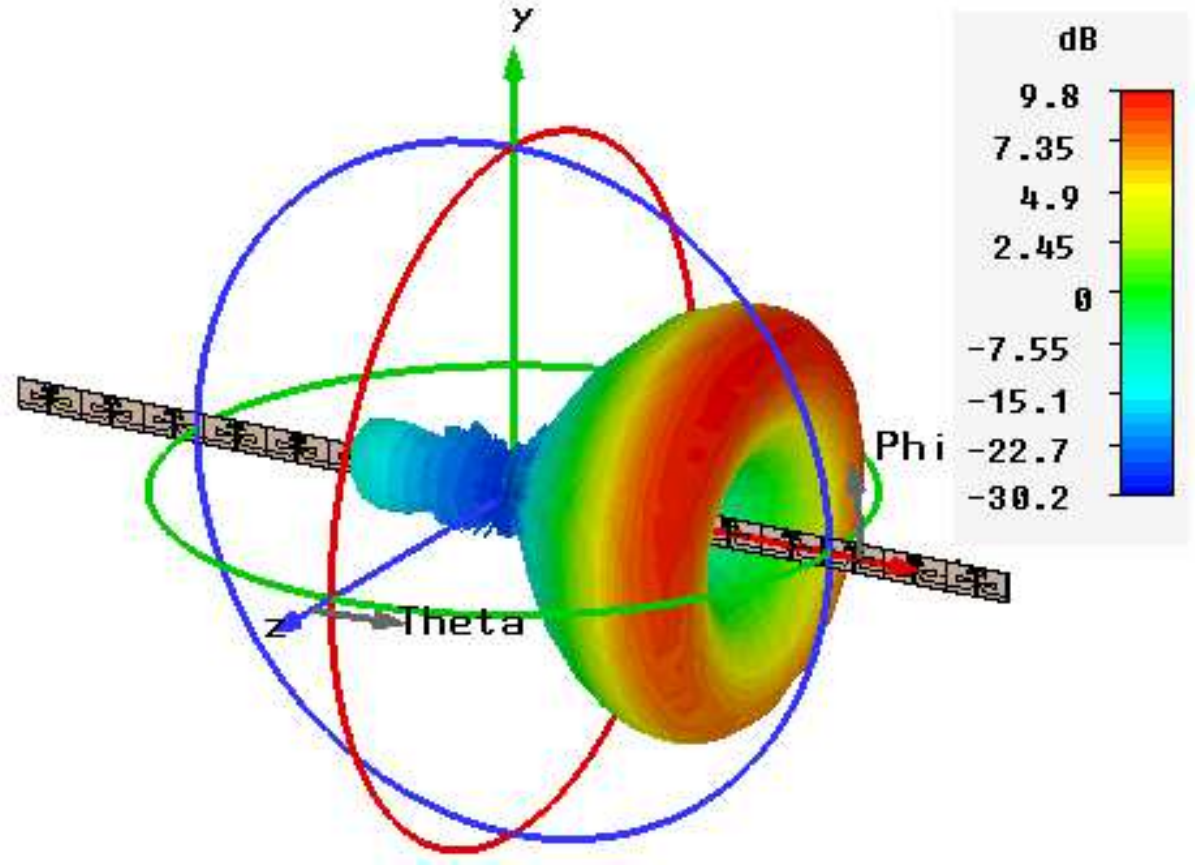}}
			\subfigure[ Front view of radiation synthesis at $40 ^\circ$]{\label{WLL_F_40_2}\includegraphics[width=5.8cm]{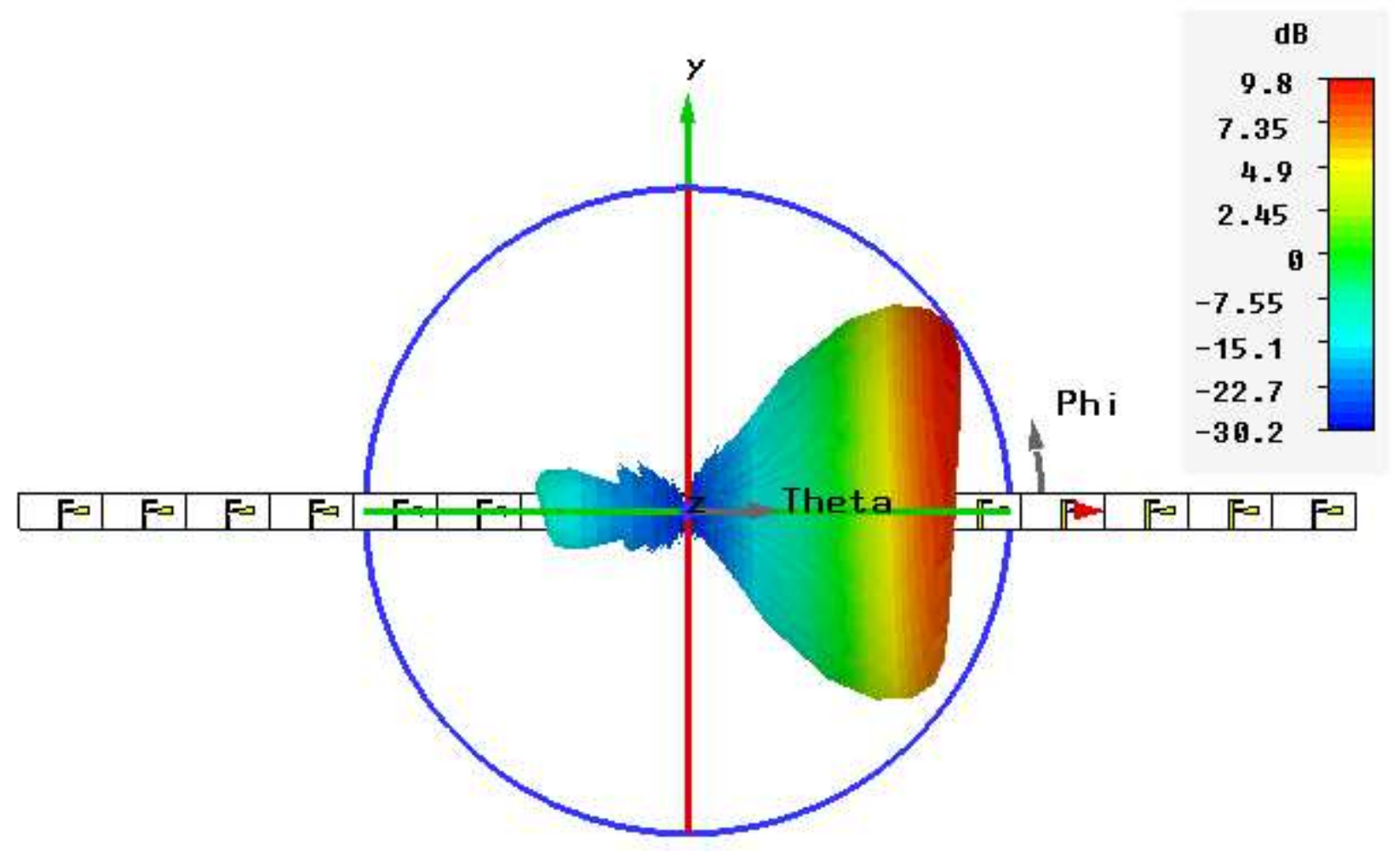}}
    \subfigure[ Cartesian representation of radiation synthesis at $40 ^\circ$]{\label{WLL_F_40_3}\includegraphics[width=7.1cm]{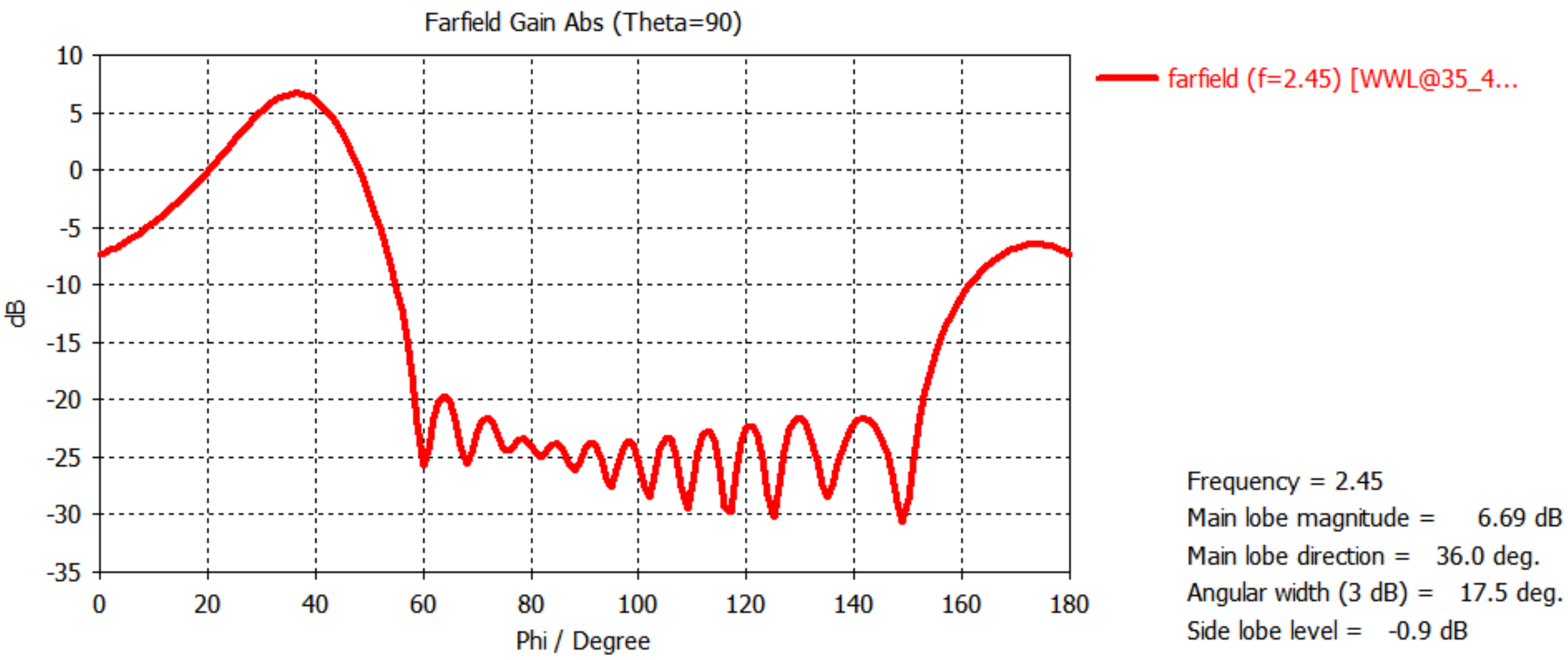}}
		
		
		
				\subfigure[ Perspective view of radiation synthesis at $70 ^\circ$]{\label{WLL_F_70_1}\includegraphics[width=4.75cm]{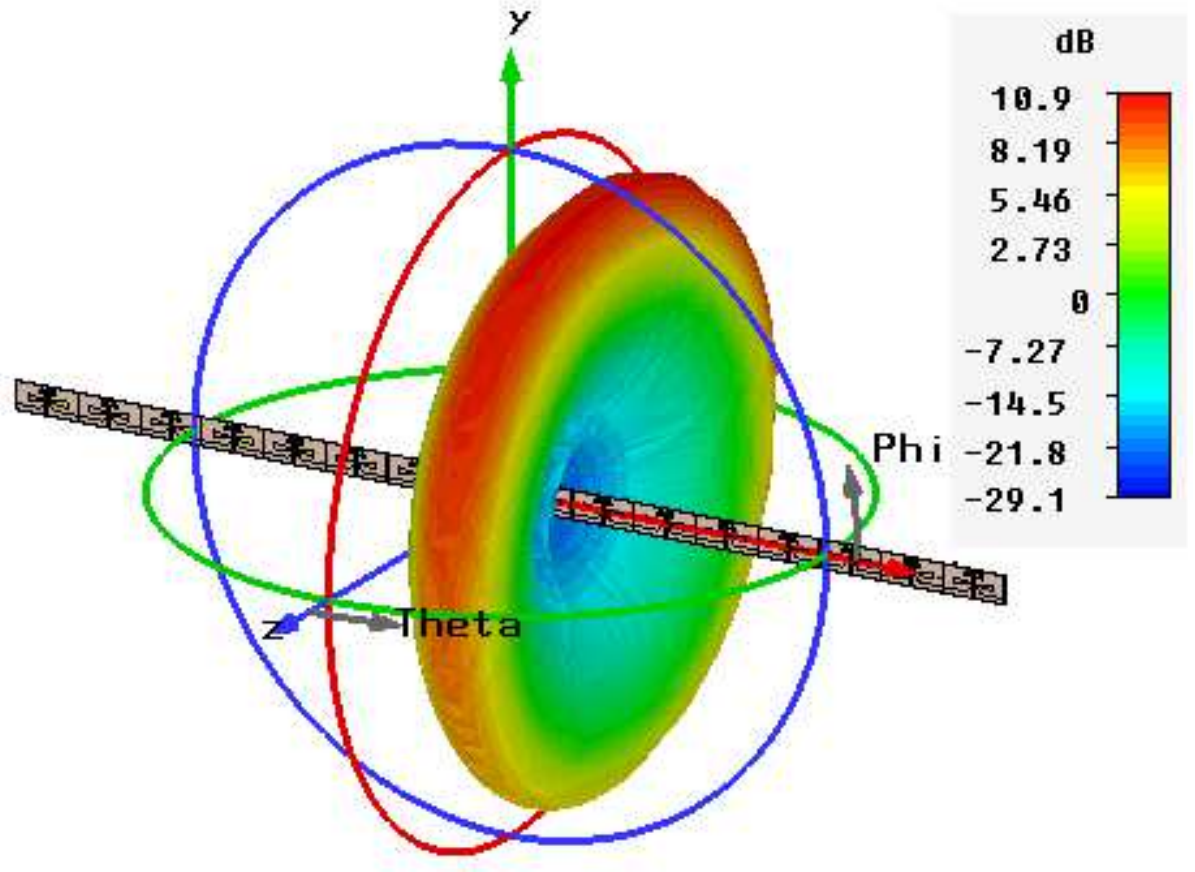}}
			\subfigure[ Front view of radiation synthesis at $70 ^\circ$]{\label{WLL_F_70_2}\includegraphics[width=5.8cm]{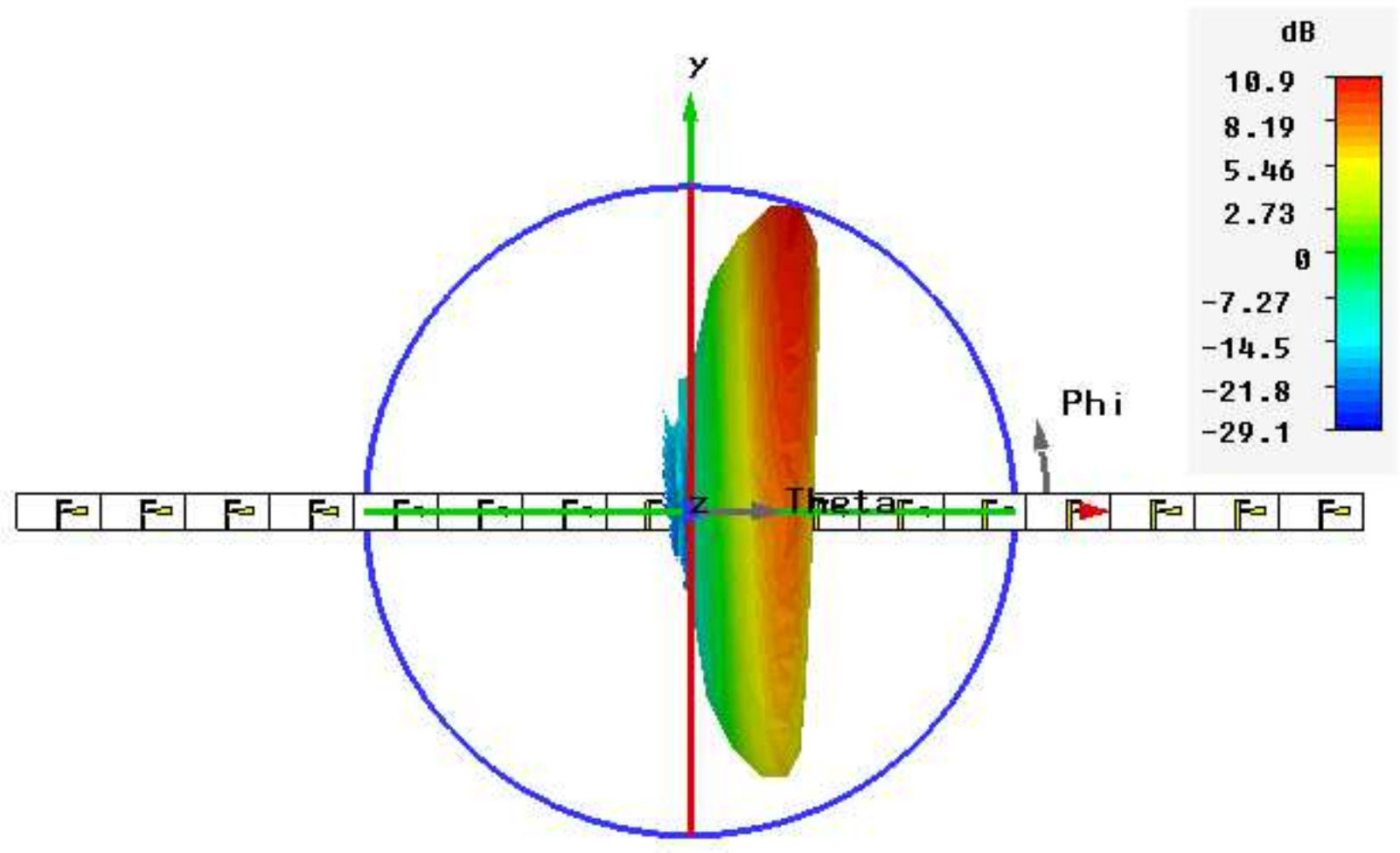}}
    \subfigure[ Cartesian representation of radiation synthesis at $70 ^\circ$]{\label{WLL_F_70_3}\includegraphics[width=7.1cm]{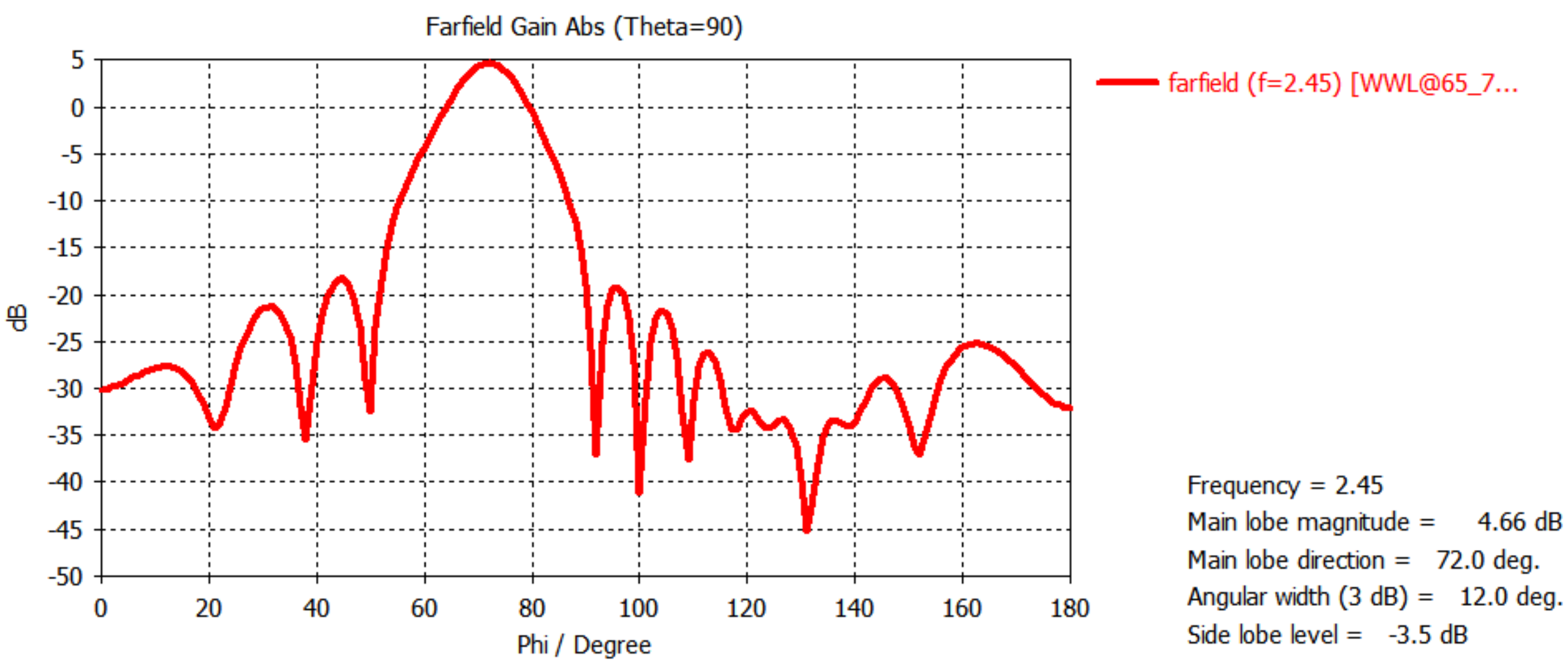}}
		


		%
		
				\subfigure[Perspective view of radiation synthesis at $120 ^\circ$]{\label{WLL_F_120_1}\includegraphics[width=4.75cm]{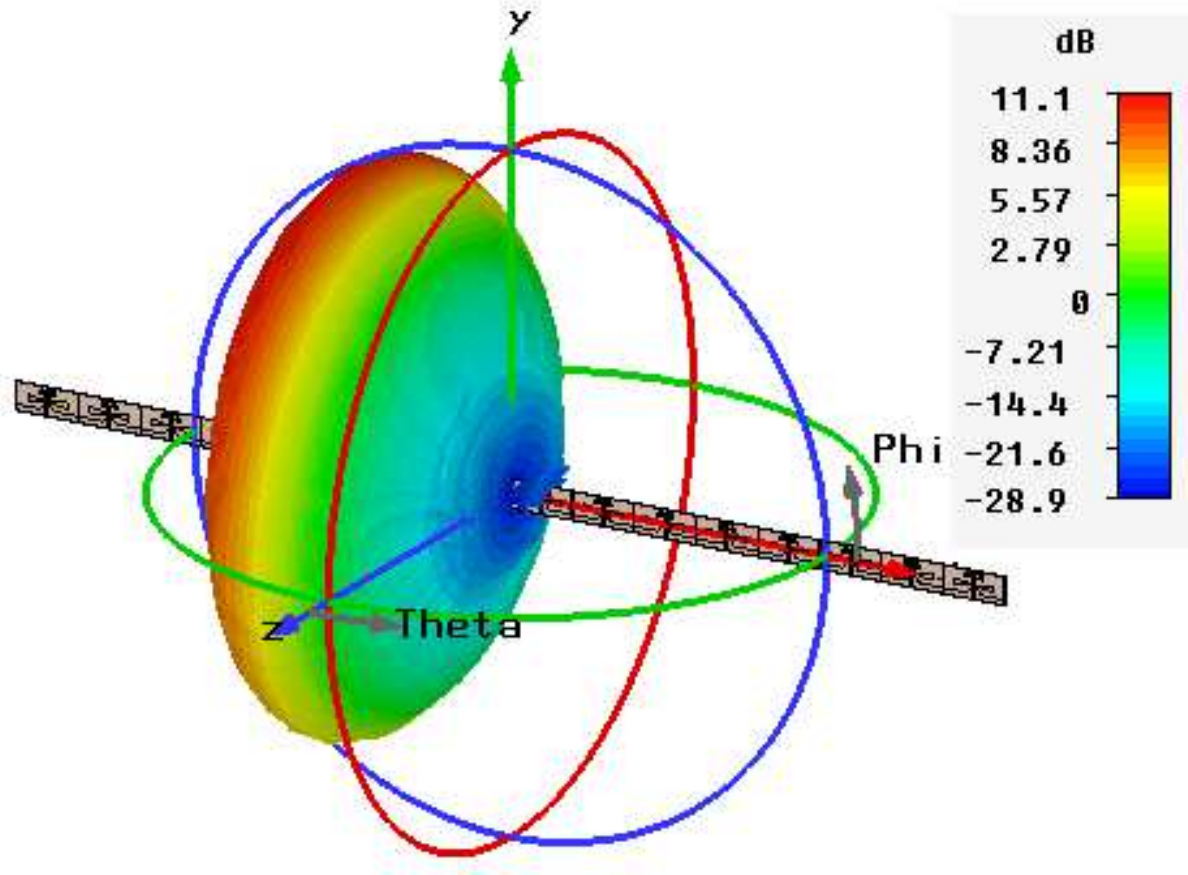}}
			\subfigure[ Front view of radiation synthesis at $120 ^\circ$]{\label{WLL_F_120_2}\includegraphics[width=5.8cm]{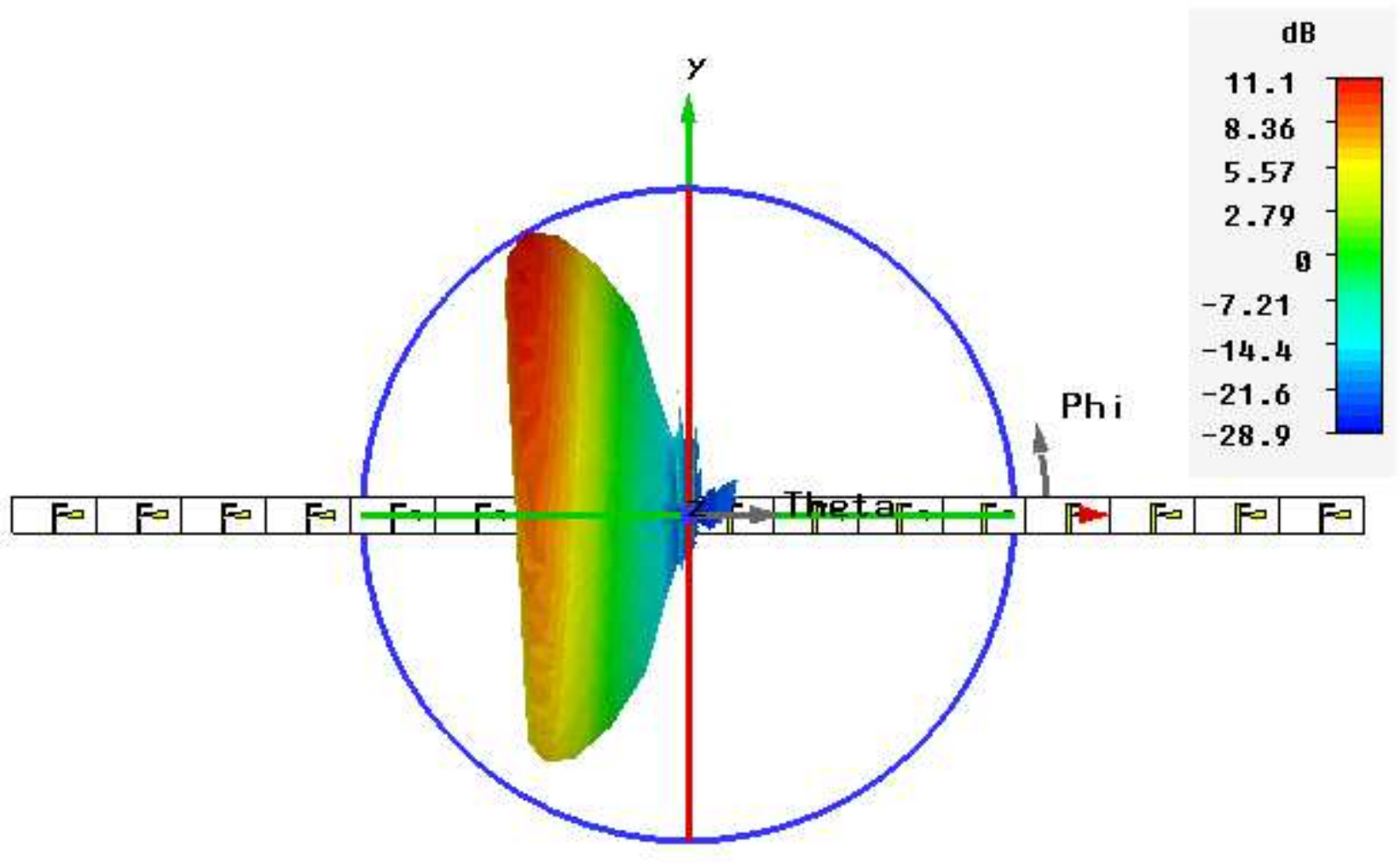}}
    \subfigure[ Cartesian representation of radiation synthesis at $120 ^\circ$]{\label{WLL_F_120_3}\includegraphics[width=7.1cm]{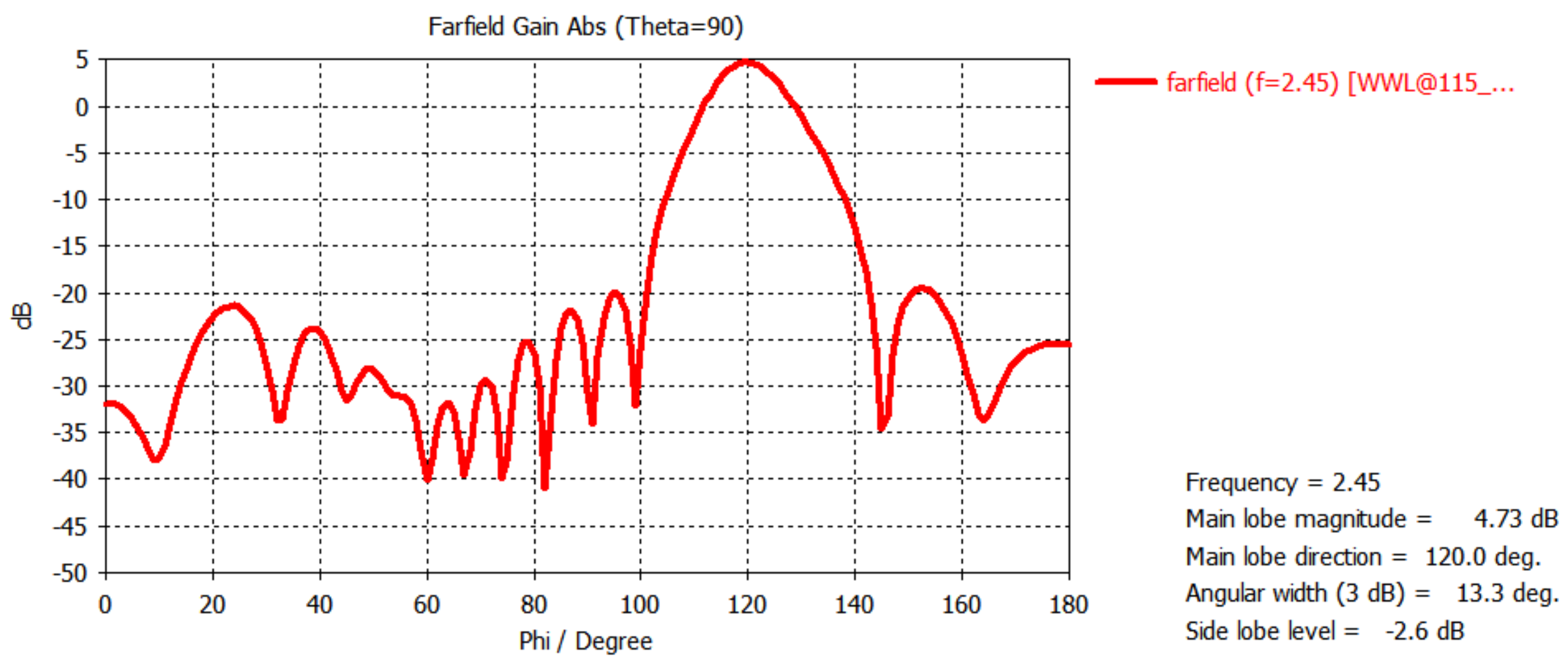}}
		
				\subfigure[ Perspective view of radiation synthesis at $130 ^\circ$]{\label{WLL_F_130_1}\includegraphics[width=4.75cm]{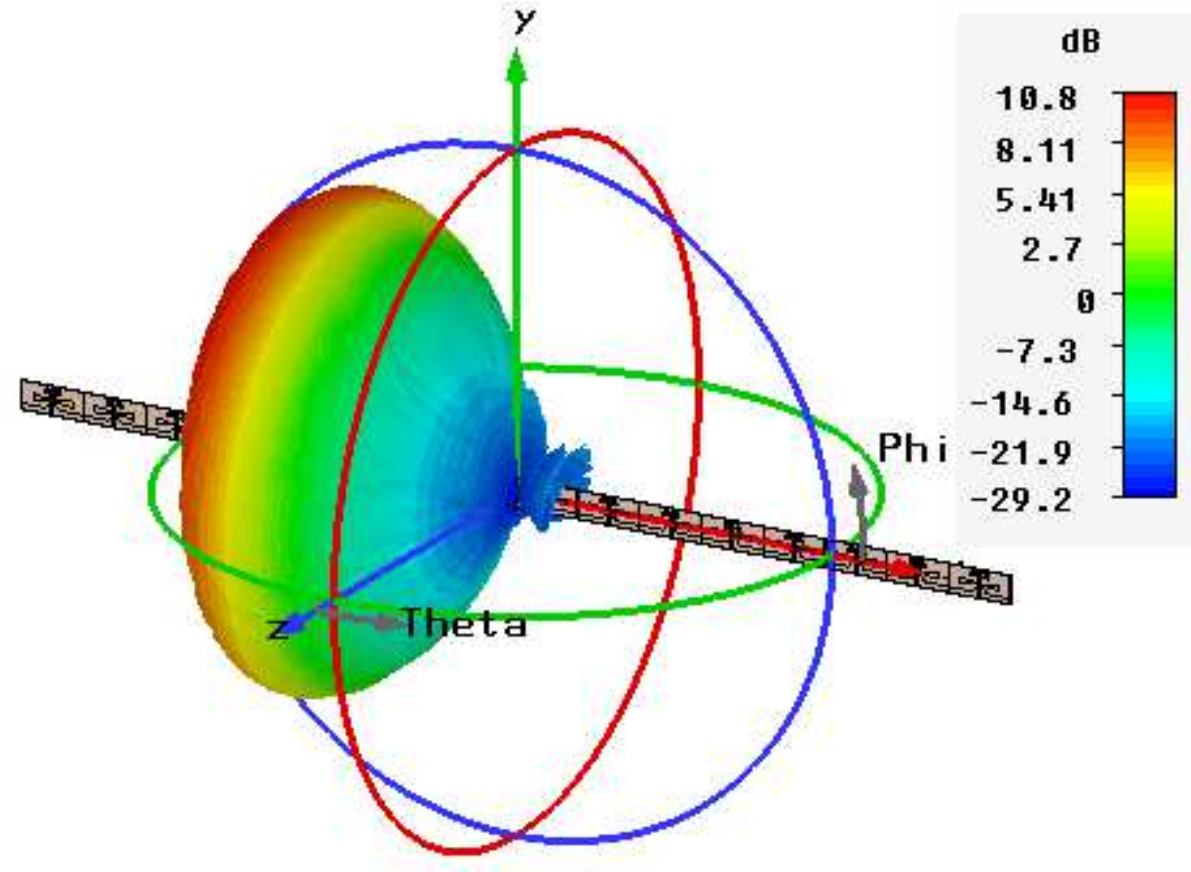}}
			\subfigure[ Front view of radiation synthesis at $130 ^\circ$]{\label{WLL_F_130_2}\includegraphics[width=5.8cm]{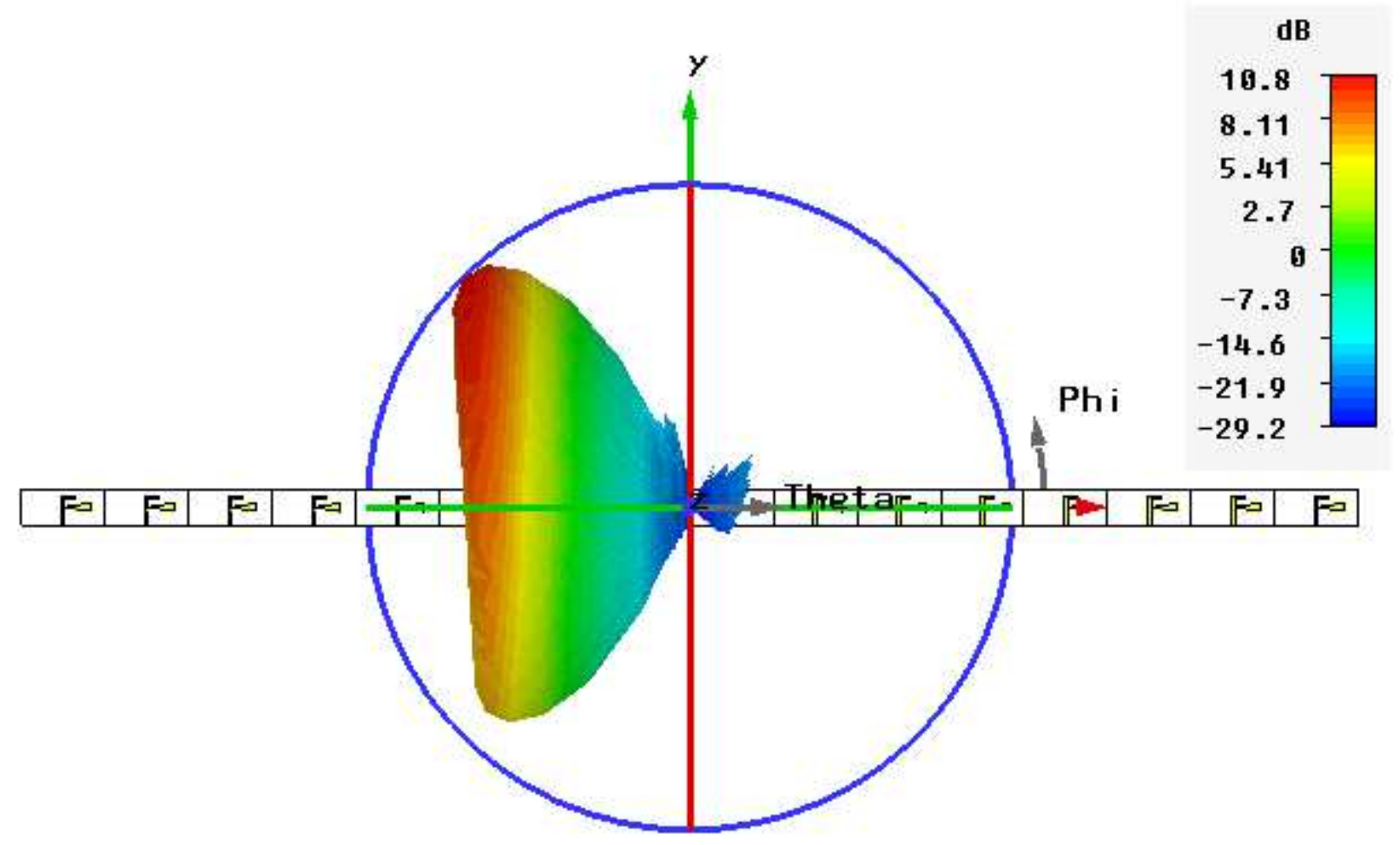}}
    \subfigure[ Cartesian representation of radiation synthesis at $130 ^\circ$]{\label{WLL_F_130_3}\includegraphics[width=7.1cm]{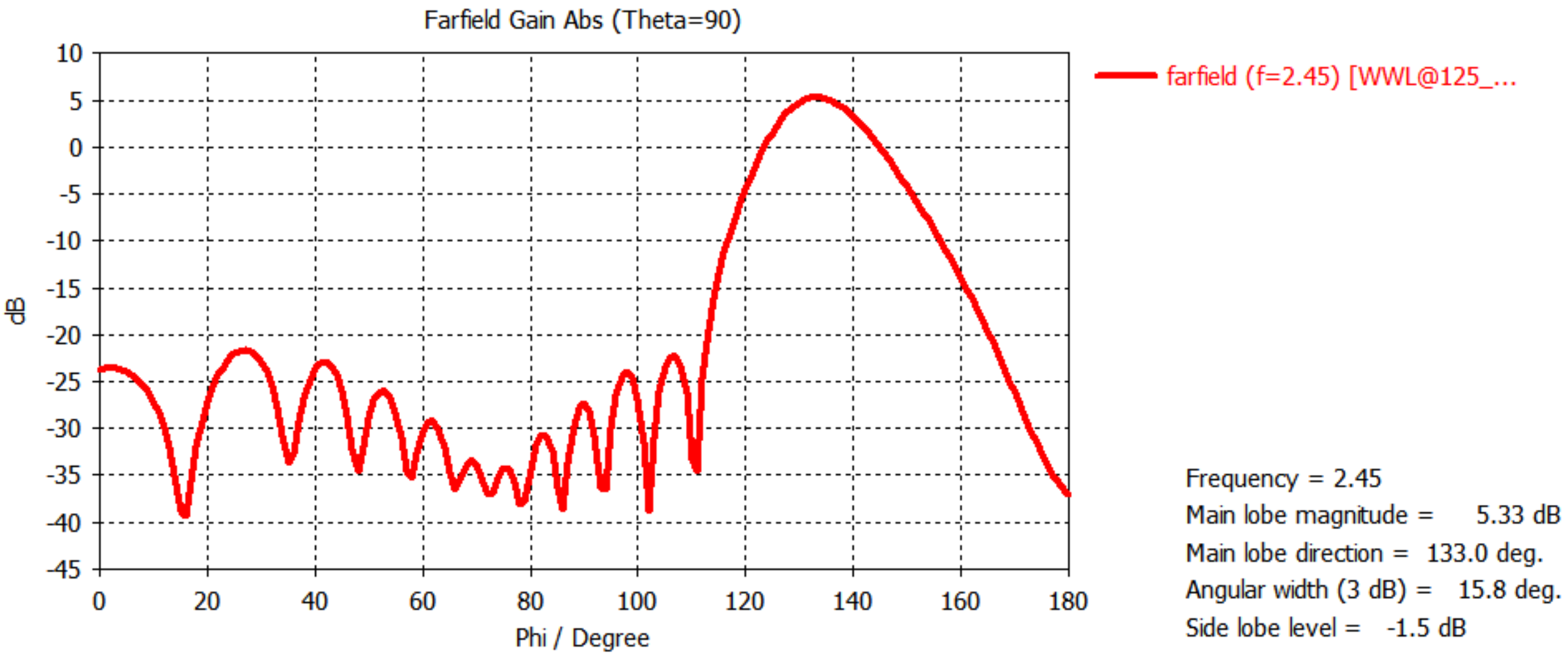}}
		

    \caption{$3D$ radiation synthesis for linear antenna array with $16$ elements using Hybrid Fourier-Woodward-Lawson-Neural Networks method at $2.45$ GHz.} 
    \label{figure_WWL_F} 
\end{figure*}

Our antenna is simulated using CST Microwave Studio. At $f=2.45$ GHz, Fig. \ref{diag_ray2} give the $3D$ omnidirectional radiation pattern with a simulated gain more than $5$ dB linearly polarized field. In this section as shown in Fig. \ref{figure_WWL_F}, an antenna array with $16$ elements is designed. For the proposed linear antenna array system, the dimensions are same as that of the single bowtie antenna shown in Fig. \ref{antenna} and the distance between the antennas is taken as $\frac{\lambda }{2}=61 mm$. Based on Fourier-Woodward-Lawson methods and NN training, TABLE \ref{tab1} and TABLE \ref{tab3} are generated; these values ​​are used as an optimal weights of the linear antenna array elements giving radiation pattern with minimum sidelobe level (SLL) and hence ameliorating the antenna array performance.

By applying optimization techniques, Fig. \ref{figure_WWL_F} illustrates many $3D$ radiation pattern of our linear antenna array (at $40^{\circ}$, $70^{\circ}$, $120^{\circ}$ and $130^{\circ}$); all these Figures with perspective view (\ref{WLL_F_40_1}, \ref{WLL_F_70_1}, \ref{WLL_F_120_1}, \ref{WLL_F_130_1}) and front view (\ref{WLL_F_40_2}, \ref{WLL_F_70_2}, \ref{WLL_F_120_2} and \ref{WLL_F_130_2}) show more the analysis of radiation patterns with these ​​optimized weighting values given by our hybrid Fourier-Woodward Lawson method using NN training. We have also different simulation results with cartesian representation of radiation pattern for $16$ antennas using the same hybrid method at $2.45$ GHz (\ref{WLL_F_40_3}, \ref{WLL_F_70_3}, \ref{WLL_F_120_3} and \ref{WLL_F_130_3}), it is clear that the criteria of SLL reduction is fulfilled. All these numerical results with all figures show the excellent phase control capability for beam pattern synthesis.

This paper illustrate the development and implementation of a new model for radiation pattern sythesis using hybrid Fourier-Woodward-Lawson-Neural Networks method for MIMO Applications to meet the desired specifications. In this paper, we succeeded to expose the performance of a hybrid optimization method in order to determine how well the optimized radiation pattern with reduced interference is suitable for a reliable MIMO wireless communication system.

\section{CONCLUSION}

In this paper, we have suggested a hybrid synthesis technique to design linear antenna array using hybrid Woodward-Lawson-Neural Networks method in order to understand and underline how well the system is appropriate to MIMO applications. To get good results, the parameters to be controlled were the antennas' excitations. This paper aims at determining radiation pattern parameters of a uniform linear antenna array. We also discussed the modeling and optimization of the synthesis problem for the antenna using Fourier method, Woodward Lawson approach, and NN training. Obtained results are encouraging enough and demonstrate that there is a clear concordance between the expected smart antennas in the specification and the synthesized ones.

\bibliographystyle{IEEEtran}
\bibliography{Biblio}

\section*{NOTICE}
\copyright~2017 IEEE. Personal use of this material is permitted. Permission from IEEE must be obtained for all other uses, in any current or future media, including reprinting/republishing this material for advertising or promotional purposes, creating new collective works, for resale or redistribution to servers or lists, or reuse of any copyrighted component of this work in other works.
\end{document}